\documentclass[a4paper,11pt]{article}
\pdfoutput=1
\usepackage{jheppub}
\usepackage[T1]{fontenc}
\usepackage{changes}

\title{\boldmath A multi-component SIMP model with $U(1)_X \rightarrow Z_2 \times Z_3$}

\author[a]{Soo-Min Choi,}
\author[b]{Jinsu Kim,}
\author[c]{Pyungwon Ko,}
\author[d]{and Jinmian Li}

\affiliation[a]{
	Physik Department T31, 
	James-Franck-Stra\ss e 1, 
	Technische Universit\"at M\"unchen, 
	D-85748 Garching, Germany}
\affiliation[b]{
	Theoretical Physics Department,
	CERN,
	1211 Geneva 23, Switzerland}
\affiliation[c]{
	School of Physics,
	Korea Institute for Advanced Study,
	Seoul 02455, Korea}
\affiliation[d]{
	College of Physics, 
	Sichuan University, 
	Chengdu 610065, China}

\emailAdd{soo-min.choi@tum.de}
\emailAdd{jinsu.kim@cern.ch}
\emailAdd{pko@kias.re.kr}
\emailAdd{jmli@scu.edu.cn}

\preprint{
TUM-HEP-1319/21\\
CERN-TH-2021-030\\
KIAS-P21008
}

\abstract{
Multi-component dark matter scenarios are studied in the model with $U(1)_X$ dark gauge symmetry that is broken into its product subgroup $Z_2 \times Z_3$  \'{a} la Krauss-Wilczek mechanism.
In this setup, there exist two types of dark matter fields, $X$ and $Y$, distinguished by different $Z_2 \times Z_3$ charges. The real and imaginary parts of the $Z_2$-charged field, $X_R$ and $X_I$, get different masses from the $U(1)_X$ symmetry breaking. 
The field $Y$, which is another dark matter candidate due to the unbroken $Z_3$ symmetry, belongs to the Strongly Interacting Massive Particle (SIMP)-type dark matter. 
Both $X_I$ and $X_R$ may contribute to $Y$'s $3\rightarrow 2$ annihilation processes, opening a new class of SIMP models with a local dark gauge symmetry.
Depending on the mass difference between $X_I$ and $X_R$, we have either two-component or three-component dark matter scenarios.
In particular two- or three-component SIMP scenarios can be realised not only for small mass difference between $X$ and $Y$, but also for large mass hierarchy between them, which is a new and unique feature of the present model.
We consider both theoretical and experimental constraints, and present four case studies of the multi-component dark matter scenarios.
}

\begin{document} 
\maketitle
\flushbottom

\section{Introduction}
\label{sec:intro}
For a few decades or so, the Weakly Interacting Massive Particle (WIMP) paradigm has been one of the mainstreams of dark matter (DM) physics (see, for example, Refs. \cite{Arcadi:2017kky,Roszkowski:2017nbc} for a recent review).
Within the WIMP scenarios, DM mass could be of the electroweak scale and its interactions with the Standard Model (SM) particles have the weak-interaction strength, and thermal relic density could be realised by freeze-out mechanism.
One would have anticipated to observe DM particle by (in)direct detections and at colliders. However the null results from the LHC searches for new particles and direct detections of DM with mass in the electroweak scale lead us to consider more seriously new paradigms beyond the WIMP paradigm.

The Strongly Interacting Massive Particle (SIMP) paradigm \cite{Hochberg:2014dra,Hochberg:2014kqa}, where the DM relic density is dominantly determined by $3\rightarrow 2$ processes in the dark sector, is one of such 
new paradigms and gained a great attention during the last several years; see Refs.~\cite{Bernal:2015bla,Lee:2015gsa,Choi:2015bya,Hochberg:2015vrg,Choi:2016hid,Choi:2016tkj,Dey:2016qgf,Choi:2017mkk,Choi:2017zww,Choi:2018iit,Hochberg:2018vdo,Dey:2018yjt,Choi:2019zeb,Maity:2019vbo,Choi:2020ara,Katz:2020ywn} for studies on various aspects of SIMP scenarios.
DM in the SIMP scenarios typically has a strong coupling and small mass scale (typically $10$ MeV--$1$ GeV) \cite{Hochberg:2014dra}. 
Thus the SIMP DM scenario is well-suited for explaining the small-scale problems of collision-less vanilla cold-DM paradigm such as the core-cusp problem and too-big-to-fail problem; 
see e.g. Ref.~\cite{Tulin:2017ara}.

As of now, parameter spaces for most of the single-component DM models are being tightly constrained by various experiments.  Single-component DM models are simple for theoretical ideas on the model building, phenomenological analyses, and comparison with various experimental data.  
However, there is no compelling reason that the DM density of the current Universe is composed of only one component. For example the visible sector of the Universe contributes only $\sim 5 \%$ to the energy density of the current Universe, and it has a number of stable or long-lived particles; electron, proton, photon, and three species of active light neutrinos.  
Presently the energy density of the dark Universe in the form of matter is about five times more than the visible Universe.  Therefore it would be an oversimplification to assume that the dark sector of the Universe consists of a single-component DM. 
Even if the dark sector of the current Universe is dominated by a single-component DM, there could be more DM species in the earlier Universe that would modify the evolution of the early Universe. They may leave some footprints that can be observed in the current Universe, e.g. on Cosmic Microwave Background or matter power spectra. 

One important feature that has to be taken into account seriously in DM physics is that DM particle should be absolutely stable or its lifetime should be much longer than the age of the Universe. 
Usually the DM stability is supposed to be guaranteed by some dark symmetries such as $Z_2$. If this dark symmetry is a global symmetry, it will be violated by gravity effects.
For example, let us consider a real scalar DM model with dark $Z_2$ symmetry under which the real scalar DM field $S$ is $Z_2$-odd, namely transforming  like $S\rightarrow -S$, whereas all the SM fields are $Z_2$-even. Then the DM $S$ will be stable at renormalizable level.
However there will be $Z_2$-breaking dimension-5 operators such as
\begin{align*}
\frac{1}{\Lambda} S ( H^\dagger H)^2 \,,\quad
\frac{1}{\Lambda} S F_{\mu\nu} F^{\mu\nu} \,,\quad
{\it etc.}
\end{align*}
Then $S$ will decay too fast to be a good cold-DM candidate even for $\Lambda\sim M_{\rm Pl}$ where $M_{\rm Pl}$ is the Planck mass, unless its mass is very light $m_S \lesssim \mathcal{O}(1)$ eV. The $Z_2$ breaking from dimension-6 operators will not be harmful at all, since one can make DM lifetime long enough for $\Lambda \ll M_{\rm Pl}$, like the proton lifetime in the SM with $B$- and $L$-violating dimension-6 operators. 
One can evade this problem of dimension-5 operators that violate dark global symmetry by promoting the dark global symmetry to a local dark gauge symmetry. 
In quantum field theory, these could be achieved by assuming that DM carries a dark charge that is either exactly conserved or approximately conserved. Finally the symmetry associated with the exactly conserved charge is implemented into a local gauge symmetry \cite{Baek:2013qwa} (see also Sec. IV B and C of Ref. \cite{Ko:2014nha} for more discussion on this issue).  
Alternatively, the approximately conserved dark charge may be an accidental symmetry of an underlying dark gauge theory, like the $U(1)$ baryon number in the SM being an accidental symmetry of the SM and being broken only by dimension-6 operators.
Thus, for either absolutely stable or long-lived DM, we end up with a local gauge theory \footnote{
One loophole in this argument is that the DM particle can be long-lived enough if it has a very small mass so that its decay channels are kinematically limited. Axion or light sterile neutrinos are good examples for such a case.
}.  
This strategy was the cornerstone of the SM of particle physics, and it could be extended to DM model building. 
In DM models with a dark gauge symmetry, there are dark gauge bosons and dark Higgs bosons in addition to the DM particles with various dark charges. 
Furthermore, there could be more than one generation of DM particles, as in the SM.
There could be more than one stable or long-lived particles in the dark sector in general.

Therefore, from theoretical point of view, it may be more natural to have a multi-component DM scenario once the symmetry that is responsible for the DM stability gets enlarged, e.g. $Z_N$ with $N\geq 4$, a product of two symmetry groups such as $Z_2 \times Z_2^\prime$, topologically stable DM, or dark gauge group with larger than rank-1, {\it etc.}
Recent studies on the multi-component DM scenarios within WIMP scenarios include Refs.~\cite{Ko:2010at,Drozd:2011aa,Aoki:2012ub,Baek:2013dwa,Ko:2014bka,Bian:2014cja,Karam:2015jta,Karam:2016rsz,Aoki:2016glu,Ko:2016fcd,Bhattacharya:2016ysw,Ahmed:2017dbb,Chakraborti:2018lso,Poulin:2018kap,YaserAyazi:2018lrv,Chakraborti:2018aae,Aoki:2018gjf,Bhattacharya:2019fgs,Chen:2019pnt,Yaguna:2019cvp}.

One of the most interesting multi-component scenarios proposed so far would be the dark QCD (DQCD) with a light Nambu-Goldstone boson DM (dark pion DM).  
One can consider both WIMP \cite{Hur:2007uz,Ko:2008ug,Ko:2009zz,Bai:2010qg,Ko:2010rj,Hur:2011sv,Bai:2013xga,Hatanaka:2016rek} and SIMP scenarios \cite{Hochberg:2014kqa,Hochberg:2015vrg,Choi:2018iit,Katz:2020ywn} in DQCD models .   
In the multi-component SIMP models proposed so far in the literature, various DM components are related with each other.  For example, dark pion DM with different dark flavours in DQCD models are related with each other by a dark flavour symmetry.  
They are not independent with each other when we consider their origin, and their masses cannot be arbitrarily different. In the case of the dark pion DM in DQCD, for example, one assumes that the dark pions are pseudo-Nambu-Goldstone bosons.
This picture requires that $m_{q_i} + m_{q_j} \ll \Lambda_{\rm dc}$, where $\Lambda_{\rm dc}$ is the dark confinement scale, which is nothing but the good old partially-conserved-axialvector-current (PCAC) condition, in order that dark pion DMs are good Nambu-Goldstone bosons. 
On the other hand, dark pion DM mass is given by $m_\pi^2 \sim m_q \Lambda_{\rm dc}$, which is bounded from above by $\Lambda^2_{\rm dc}$. Therefore, one cannot imagine a wide separation in the dark pion DM masses with different flavours. 
They are bounded by $\Delta m_\pi^2 \sim \Lambda^2_{\rm dc}$.
In other words, we can say that they are not genuine multi-component SIMP models where arbitrary masses are allowed, including the large mass hierarchy.  
For genuine multi-component DM models, different DM components should have different masses (and different lifetimes in the case of a decaying DM), different (dark) charges, and carry even different spins.  

In this paper we introduce two independent fields $X$ and $Y$ with different dark charges (and even different spins) from the beginning, and show that two- or three-component SIMP scenarios can be realised.  
Therefore the model we present in the following could be taken as a real multi-component SIMP scenario. Our study is not based on the effective field theory approach. The model constructed in this work is renormalizable and gauge invariant under the SM and the dark gauge symmetry groups, and thus a UV-complete one.  
Our conclusion can thus be taken as reliable in the entire energy range considered.

At this point, it is worthwhile to remind ourselves that WIMP and SIMP are about mechanisms to reproduce observed DM relic density in a given DM model, and not about the underlying particle physics models for DM. Within the same DM model, different mechanisms can be realised depending on the parameter values and mass scales. Likewise, either single-component or multi-component WIMP or SIMP scenarios can be realised within the same DM model, depending on the parameter values and mass scales of various particles in the model.
The model discussed in the following shows how rich the dark sector can be, depending on the parameter spaces (couplings and mass scales), and how DM particles are stabilised ($U(1)$ dark charge assignments in our case). The multi-component SIMP scenarios are realised only in a particular corner of the parameter space.  
In a generic parameter space, we expect that single-component SIMP scenarios or WIMP scenarios would be realised, but these two options shall not be discussed in detail, since they are not qualitatively different from the literature of the SIMP/WIMP scenarios, and thus are not the main interest of this work.

In this work we consider a multi-component DM model with $U(1)_X$ dark gauge symmetry broken into its product subgroup $Z_2 \times Z_3$ \'{a} la Krauss-Wilczek mechanism \cite{Krauss:1988zc}.
There exist two types of DM fields, $X$ and $Y$, which are distinguished by different $U(1)_X$ charges with the following symmetry breaking pattern: $ U(1)_X \rightarrow Z_2 \times Z_3$.
The stability of $X$ is guaranteed by the unbroken $Z_2$ symmetry, while the unbroken $Z_3$ symmetry ensures the stability of $Y$. 
Then we can realise a multi-component SIMP scenario through various DM number-changing scattering processes such as $YYY\rightarrow XX$. 
Note that $X$ can be either scalar or fermion, whereas $Y$ should be scalar.
The relic densities of different DM fields are determined through interactions between the dark fields. For the $Z_2$-charged field $X$, number-changing $2\rightarrow2$ processes are responsible for $X$'s relic density. On the other hand, the relic density of the $Z_3$-charged field $Y$ is given by number-changing $3\rightarrow2$ processes. As the processes responsible for the relic densities of the DM fields occur within the dark sector and both $X$ and $Y$ participate in the number-changing processes, we dub our model a multi-component SIMP DM scenario.
The field $X$ receives different masses for its real and imaginary parts from the vacuum expectation value (VEV) of the dark Higgs, breaking the $U(1)_X$ symmetry. Due to the interactions between $X$ and $Y$, both $X_I$ and $X_R$ may contribute to $Y$'s $3\rightarrow 2$ processes. It opens a new class of SIMP models.
Depending on the mass gap between $X_I$ and $X_R$, two-component or three-component DM scenarios can be realised.
One of the most distinctive features of this model is that the different DM fields $X$ and $Y$ may have vastly different masses, while significantly contributing to the total DM relic density today.
This is in sharp contrast to most of the studies on multi-component SIMP where each component could coexist when the mass difference is small enough \cite{Choi:2016hid,Hochberg:2018vdo,Katz:2020ywn}. For completeness we consider both cases with large mass hierarchy and small mass difference.

It should be emphasised that the present work, which considers multi-component SIMP scenarios with a large mass hierarchy, is mostly motivated by theoretical reasons, as a kind of existence proof of such a case within a theoretically well-defined DM model. Still there is an interesting phenomenological consequence.  
The coupling used to maintain the kinetic equilibrium between the DM sector and the SM sector may also be used to detect the boosted $Y$ (from $X$ annihilation) in DM and neutrino detectors. 
The boosted DM signal is one of the most important predictions of the multi-component DM scenarios with large mass hierarchy.

The paper is organised as follows. We first describe our model setup in Section \ref{sec:model}, where we set the notations, explain the $U(1)_X$ symmetry breaking, and identify DM candidates. In Section \ref{sec:dmpheno}, we study phenomenology of the multi-component DM scenarios. The evolution of the DM system is described by four coupled Boltzmann equations, each for $Y$, $X_I$, $X_R$, and $Z^\prime$. We numerically solve the Boltzmann equations and present two benchmark cases for the two-component scenario and two benchmark cases for the three-component scenario. In Section \ref{subsec:constraints}, we consider theoretical constraints, such as perturbativity and unitarity, as well as experimental constraints, including LHC constraints for the invisible Higgs decay, the DM direct detection, and $Z^\prime$ searches. One of the crucial constraints for a SIMP-type DM is the sufficient release of its kinetic energy. We address this issue in detail in Section \ref{subsec:constraints}. Taking into account all of the constraints we present our results in Sections \ref{subsec:2compnt} and \ref{subsec:3compnt}.
Finally we conclude in Section \ref{sec:conclusions}.

\section{Model}
\label{sec:model}
We consider a dark $U(1)_X$ gauge symmetry with three different types of scalars in the dark sector: two DM candidates $X$ and $Y$ plus the dark Higgs field $\phi$.
We assign the $U(1)$ charges of these fields as follows~\footnote{
We can also consider another possibility with spin-1/2 dark fermion $\psi$ with $q_\psi = 1/2$, instead of dark scalar $X$. We shall not pursue this case in detail in this paper, since the qualitative features of multi-component SIMP scenarios would be the same.}:
\begin{align}
( q_X, q_Y, q_\phi ) = (1/2,1/3,1)\,.
\label{eqn:U1X-charge}
\end{align}
Then, the gauge-invariant renormalizable Lagrangian of this model is given by
\begin{align}
\mathcal{L} &=
\mathcal{L}_{\overline{{\rm SM}}}
-\frac{1}{4}\hat{X}_{\mu\nu}\hat{X}^{\mu\nu}
-\frac{\sin\epsilon}{2}\hat{X}_{\mu\nu}\hat{B}^{\mu\nu}
+ |D_{\mu} H|^{2} + |D_\mu X|^{2} + |D_\mu Y|^{2} + |D_\mu \phi|^{2}
-V
\,,
\end{align}
where $\mathcal{L}_{\overline{{\rm SM}}}$ is the SM Lagrangian without the Higgs sector, $H$ is the SM Higgs field, and $\epsilon$ is the $U(1)_Y$--$U(1)_X$ gauge kinetic mixing parameter. 
The covariant derivative of the dark sector is defined as $D_\mu \equiv \partial_\mu - iq_ig_X\hat{X}_\mu$, with $g_X$ being the dark gauge coupling and the $U(1)_X$-charge assignments of dark fields $q_i$ $(i=X,Y,\phi)$ are defined in Eq.~\eqref{eqn:U1X-charge}. 

The scalar potential is given by
\begin{align}\label{eqn:potential}
V &= m_{X}^{2}|X|^{2} + m_{Y}^{2}|Y|^{2}
+\lambda_{\phi}\left(
|\phi|^{2} - \frac{v_{\phi}^{2}}{2}
\right)^{2}
+\lambda\left(
|H|^{2} - \frac{v^{2}}{2}
\right)^{2}
+\lambda_{X}|X|^{4}
+\lambda_{Y}|Y|^{4}
\nonumber\\
&\quad
+\lambda_{XY}|X|^2|Y|^2
+\lambda_{X\phi}|X|^{2}\left(
|\phi|^{2} - \frac{v_{\phi}^{2}}{2}
\right)
+\lambda_{Y\phi}|Y|^{2}\left(
|\phi|^{2} - \frac{v_{\phi}^{2}}{2}
\right)
\nonumber\\
&\quad
+\lambda_{XH}|X|^{2}\left(
|H|^{2} - \frac{v^{2}}{2}
\right)
+\lambda_{YH}|Y|^{2}\left(
|H|^{2} - \frac{v^{2}}{2}
\right)
+\lambda_{\phi H}\left(
|\phi|^{2} - \frac{v_{\phi}^{2}}{2}
\right)\left(
|H|^{2} - \frac{v^{2}}{2}
\right)
\nonumber\\
&\quad
+\left[
\mu_{X\phi}X^{\dagger 2}\phi
+\lambda^{\prime}_{Y\phi}Y^{\dagger 3}\phi
+\text{h.c.}
\right]
\,,
\end{align}
where $v_\phi$ and $v$ are VEVs of the dark Higgs and the SM Higgs fields.
After $U(1)_X$ symmetry breaking by a nonzero VEV $\langle \phi \rangle = v_\phi / \sqrt{2}$, the last line of Eq.~\eqref{eqn:potential} still has $Z_2 \times Z_3$ discrete gauge symmetries~\footnote{
See also Refs. \cite{Baek:2014kna,Baek:2020owl,Kang:2021oes} and Refs. \cite{Ko:2014nha,Ko:2014loa,Guo:2015lxa,Ko:2020gdg} for scalar DM models with $U(1)$ dark gauge symmetry broken into $Z_2$ and $Z_3$, respectively.}: 
\begin{align*}
X & \rightarrow e^{i \pi} X = - X \,, 
\\
Y & \rightarrow e^{\pm i 2 \pi /3} Y \,. 
\end{align*}

We note that $\mu_{X\phi}$ has the mass dimension one.
Expanding the dark Higgs and the SM Higgs around their VEVs, $v_{\phi}$ and $v$, and writing $X = (X_{R} + i X_{I})/\sqrt{2}$, the real and the imaginary parts of $X$ get different masses from the $U(1)_X$ symmetry breaking. The mass splitting between $X_R$ and $X_I$ is generated by the $\mu_{X\phi}$ term \cite{Baek:2014kna}:
\begin{align}
m_{X_{I}}^{2} &=
m_{X}^{2} - \sqrt{2}\mu_{X\phi}v_{\phi}
\,,\\
m_{X_{R}}^{2} &=
m_{X}^{2} + \sqrt{2}\mu_{X\phi}v_{\phi}
\,.
\end{align}
Taking $\mu_{X\phi}>0$ without loss of generality, we see that $X_I$ is always lighter than $X_R$.

The mass-squared matrix of the Higgs fields is given by
\begin{align}
M^2 = \left(
\begin{array}{cc}
m_{hh}^{2} & m_{hh^\prime}^{2} \\
m_{hh^\prime}^{2} & m_{h^\prime h^\prime}^{2}
\end{array}
\right)\,,
\end{align}
with~\footnote{
We denote the physical SM Higgs field and the physical dark Higgs field by $h$ and $h^\prime$, respectively.
}
\begin{align}
m_{hh}^{2} = 2\lambda v^{2}
\,,\quad
m_{h^\prime h^\prime}^{2} = 2\lambda_{\phi} v_{\phi}^{2}
\,,\quad
m_{hh^\prime}^{2} = \lambda_{\phi H}vv_{\phi}
\,.
\end{align}
The mass eigenvalues are then obtained as follows:
\begin{align}
m_{1,2}^{2} = \frac{1}{2}\left[
(m_{hh}^{2} + m_{h^\prime h^\prime}^{2})
\mp
\sqrt{(m_{hh}^{2} - m_{h^\prime h^\prime}^{2})^{2}
+ 4m_{hh^\prime}^{4}}
\right]\,.
\end{align}
The mass eigenstates, which we denote by $h_{1,2}$, are related to the interaction eigenstates as
\begin{eqnarray}
\left(
\begin{array}{c}
h \\ h^\prime
\end{array}
\right) = \left(
\begin{array}{cc}
\cos\alpha & -\sin\alpha \\
\sin\alpha & \cos\alpha
\end{array}
\right)\left(
\begin{array}{c}
h_2 \\ h_1
\end{array}
\right)\,,
\end{eqnarray}
with the mixing angle
\begin{eqnarray}
\alpha = \frac{1}{2}\tan^{-1}\left[
\frac{2m_{hh^\prime}^{2}}{m_{hh}^{2} - m_{h^\prime h^\prime}^{2}}
\right]\,.
\end{eqnarray}

With the $U(1)_Y$--$U(1)_X$ gauge kinetic mixing, the mass matrix of the gauge boson can be diagonalised as follows. The kinetic mixing can be removed in terms of $\tilde{B}_\mu$ and $\tilde{X}_\mu$ defined as \cite{Babu:1997st},
\begin{eqnarray}
\left(
\begin{array}{c}
\hat{B}_{\mu} \\ \hat{X}_{\mu}
\end{array}
\right) = \left(
\begin{array}{cc}
1 & -\tan\epsilon \\
0 & 1/\cos\epsilon
\end{array}
\right)\left(
\begin{array}{c}
\tilde{B}_{\mu} \\ \tilde{X}_{\mu}
\end{array}
\right)\,,\qquad
\hat{W}_\mu = \tilde{W}_\mu\,,
\end{eqnarray}
where we defined $\tilde{W}_\mu$ for notational consistency.
We then diagonalise the mass matrix for $\tilde{B}_\mu$, $\tilde{X}_\mu$, and $\tilde{W}^3_\mu$ as
\begin{align}
\left(
\begin{array}{c}
\tilde{B}_{\mu} \\ \tilde{W}^3_\mu \\ \tilde{X}_{\mu}
\end{array}
\right) = \left(
\begin{array}{ccc}
c_w & -s_w c_\zeta & s_w s_\zeta \\
s_w & c_w c_\zeta & -c_w s_\zeta \\
0 & s_\zeta & c_\zeta
\end{array}
\right)\left(
\begin{array}{c}
A_\mu \\ Z_\mu \\ Z^\prime_\mu
\end{array}
\right)\,,
\end{align}
where $c_w \equiv \cos\theta_w$, $s_w \equiv \sin\theta_w$, $c_\zeta \equiv \cos\zeta$, $s_\zeta \equiv \sin\zeta$, and $\theta_w$ is the Weinberg angle. Here $\zeta$ is defined as
\begin{eqnarray}
\tan 2\zeta \equiv
-\frac{m_{\hat{Z}}^{2}s_w \sin 2\epsilon}
{m_{\hat{X}}^2 - m_{\hat{Z}}^2 (c_\epsilon^2 - s_\epsilon^2 s_w^2)}\,,
\end{eqnarray}
where $m_{\hat{Z}} = \sqrt{(g_1^2 + g_2^2)}v/2$, $m_{\hat{X}} = g_X v_\phi$, $c_\epsilon \equiv \cos\epsilon$, and $s_\epsilon \equiv \sin\epsilon$.
Subsequently, we obtain the following relations between the original fields and the mass eigenstates:
\begin{eqnarray}
\hat{B}_{\mu} &=&
c_w A_{\mu} - (t_\epsilon s_\zeta + s_w c_\zeta)Z_\mu
+(s_w s_\zeta - t_\epsilon c_\zeta) Z_{\mu}^{\prime}
\,,\nonumber\\
\hat{X}_{\mu} &=&
\frac{s_\zeta}{c_\epsilon}Z_{\mu}
+\frac{c_\zeta}{c_\epsilon}Z_{\mu}^{\prime}
\,,\\
\hat{W}^3_{\mu} &=& s_w A_{\mu} + c_w c_\zeta Z_{\mu}
-c_w s_\zeta Z_{\mu}^{\prime} \,,\nonumber
\end{eqnarray}
with $t_\epsilon \equiv \tan \epsilon$ (similarly for $\zeta$ below).
The masses of the $Z$ and $Z^{\prime}$ gauge bosons are given by
\begin{eqnarray}
m_{Z}^{2} = m_{\hat{Z}}^{2}(1 + s_{w}t_{\zeta}t_{\epsilon})
\,,\quad
m_{Z^{\prime}}^{2} = \frac{m_{\hat{X}}^{2}}
{c_{\epsilon}^{2}(1+s_{w}t_{\zeta}t_{\epsilon})}
\,.
\end{eqnarray}
In the following we assume small mixing angles, i.e. $\alpha \ll 1 $ and $\epsilon \ll 1$, hence $h_2 \approx h$ and $h_1 \approx h^\prime$, and $m_{Z}^{2} \approx m_{\hat{Z}}^{2}$ and $m_{Z^{\prime}}^{2} \approx m_{\hat{X}}^{2}$.

In our model, due to the unbroken $Z_2 \times Z_3$ symmetry, fields with $U(1)_X$ charges of 1/2 and 1/3 will be stable, becoming DM candidates. The $X_I$ field with $q_X=1/2$ is absolutely stable and makes one of the DM components. The heavier one $X_R$ decays into $X_I, Z^{\prime(*)}$, and could be another good DM candidate, if its lifetime is long enough compared with the age of the Universe. 
When the mass splitting is so small, $m_{X_R} - m_{X_I} \ll {\rm min}\{2m_Y,m_{Z^\prime}\}$, that the $X_R$ decay channels such as $X_R \rightarrow X_I,Z^\prime$ and $X_R \rightarrow X_I,Y,Y^*$ are closed, the $X_R$ field again becomes a good DM candidate (see Appendix~\ref{apdx:XRdecay} for the three-body decay of $X_R$).
The field $Y$ with $q_Y = 1/3$ is another DM candidate because of the unbroken $Z_3$ symmetry.  
Then for example 
\begin{align*}
Y,Y,Y \rightarrow X_I, X_I  \,, \quad {\it etc.}
\end{align*}
could be possible through the dark Higgs and/or dark gauge boson exchanges, which is a new class of SIMP models based on spontaneous breaking of dark $U(1)_X \rightarrow Z_2 \times Z_3$.

We are primarily interested in the setup where the field $Y$ is the lightest among the DM candidates. To implement the SIMP scenario for $Y$, we assume that the dark Higgs and the dark gauge boson are heavier than $Y$. Otherwise, the DM will be pair-annihilated into dark Higgs/dark gauge boson, which is not of interest in this work.
Similarly, if either the mixing between the SM Higgs boson and dark Higgs or the kinetic mixing between gauge bosons of $U(1)_Y$ and $U(1)_X$ is sizeable, the DM pair-annihilating into SM particles will be important once kinematically open. We shall therefore focus on the parameter spaces that have small mixing between SM Higgs and dark Higgs, as well as small kinetic mixing in the gauge sector. We note that this setup is also useful to suppress the DM-nucleon scattering cross section, which is tightly constrained by the DM direct detection experiments. 
On the other hand, it is necessary for $Y$ to lose its kinetic energy through elastic scattering with SM particles \cite{Hochberg:2014dra,Choi:2015bya}. For this, a somewhat light $Z^\prime$ is needed for the SIMP--SM kinetic equilibrium condition as we discuss in Section~\ref{subsec:constraints}.
We focus on the case where $m_{Z^\prime} > 2m_Y$; thus $Z^\prime$ is not stable and will decay. As long as the decay is fast enough, $Z'$ cannot be a DM candidate. 
For the dark Higgs, we assume that it is much heavier than $Y$, $X_{I,R}$, and $Z^\prime$ such that the dark Higgs does not play any significant role in the dynamics.
The cross section for the process $X,X \rightarrow Y,Y^*$ features a four-point vertex~\footnote{
We note that contact interactions between $Z_2$-charged field and $Z_3$-charged field are absent at renormalizable level in the fermionic $Z_2$ DM, where the scalar field $X$ is replaced by a spin-1/2 dark fermion $\psi$.
} which, in the large-$m_{h^\prime}$ limit, is proportional to $\lambda_{XY}^2$.
The relic density of the $X$ field will be diluted through this process if $\lambda_{XY}$ is large.
When the dark Higgs is somewhat lighter, couplings $\lambda_{X\phi}$ and $\lambda_{Y\phi}$ need to be set appropriately such that the $X,X\rightarrow Y,Y^*$ is suppressed via the destructive interferences between the contact process and dark Higgs mediated process.
We first assume that $\lambda_{XY}$ is small enough such that this contact $X$-annihilation process is not efficient in the large-$m_{h^\prime}$ limit.
We then comment on the case of destructive interferences, where $\lambda_{XY}$ may take a larger value, presenting parameter sets with a lighter dark Higgs and larger dark Higgs-portal couplings.

In summary the DM candidates are $X_R$, $X_I$, and $Y$ in our setup.
Our interest is to investigate whether two- or three-component DM scenarios can be realised for the following four cases: {\it i)} $m_{X_R} \gg m_{X_I} \gg m_Y$, {\it ii)} $m_{X_R} \gg m_{X_I} \sim m_Y$, {\it iii)} $m_{X_R} \sim m_{X_I} \gg m_Y$, and {\it iv)} $m_{X_R} \sim m_{X_I} \sim m_Y$. The cases {\it i)} and {\it ii)} will be the two-component DM scenarios, while {\it iii)} and {\it iv)} correspond to the three-component scenarios.
We do not consider the inverted mass hierarchy, i.e. $m_{X_I} \lesssim m_{X_R} \ll m_Y$. On one hand, in order for the $Z_3$-charged field $Y$ to serve as a SIMP DM, the mass should be less than $\sim$50 MeV, in the large-$m_{h^\prime}$ limit, as number-changing $3\rightarrow2$ processes are dominant. On the other hand, a DM lighter than $\sim$5 MeV is heavily constrained by cosmology such as Big Bang Nucleosynthesis and Cosmic Microwave Background, as explored in Ref.~\cite{Sabti:2019mhn}. Therefore, we do not consider the inverted mass hierarchy in this work, and focus on the aforementioned cases {\it i)}--{\it iv)}.

\section{Multi-component Dark Matter Phenomenology}
\label{sec:dmpheno}

\subsection{Generalities}
Under the assumptions described in the previous section, we have four coupled Boltzmann equations for describing the evolution of the system consisting of $\{Y, X_I, X_R, Z^\prime\}$~\footnote{
Since the dark Higgs is assumed to be sufficiently heavier than the rest of the dark sector fields, it does not play any significant role in the dynamics.
}. 
Schematically the Boltzmann equation for the species $i$ is given by
\begin{align}
\dot{n}_i + 3Hn_i &=
-\langle \Gamma \rangle_{i\rightarrow j,k} \left(
n_i - n_i^{\rm eq}\frac{n_j n_k}{n_j^{\rm eq} n_k^{\rm eq}}
\right)
-\langle \Gamma \rangle_{i\rightarrow j,k,l} \left(
n_i - n_i^{\rm eq}\frac{n_j n_k n_l}{n_j^{\rm eq} n_k^{\rm eq} n_l^{\rm eq}}
\right)
\nonumber\\
&\quad
-\langle \sigma v \rangle_{i,j \rightarrow l,m}
\left(
n_i n_j - n_i^{\rm eq}n_j^{\rm eq}\frac{n_l n_m}{n_l^{\rm eq}n_m^{\rm eq}}
\right)
\nonumber\\
&\quad
-\langle \sigma v^2 \rangle_{i,j,k \rightarrow l,m} 
\left(
n_i n_j n_k - n_i^{\rm eq}n_j^{\rm eq}n_k^{\rm eq}
\frac{n_ln_m}{n_l^{\rm eq}n_m^{\rm eq}}
\right)
\,,
\end{align}
where $n_i$ ($n_i^{\rm eq}$) is the (equilibrium) number density of the species $i$, $H$ the Hubble parameter, and $\dot{} \equiv d/dt$ with $t$ being the cosmic time.

Here $\langle \sigma v \rangle$ and $\langle \sigma v^2 \rangle$ are the thermally-averaged cross sections, and $\langle \Gamma \rangle$ is the thermally-averaged decay rate.
The thermally-averaged cross section $\langle \sigma v \rangle$ for the $i,j\rightarrow l,m$ process is given by
\begin{align}
\langle \sigma v \rangle = \frac{1}{S}\frac{1}{F}\sum_{\rm spins}\frac{1}{64\pi^2 E_i E_j (E_i+E_j)}\int d\Omega |\mathbf{p}_f||\mathcal{M}|^2\,,
\end{align}
where $F$ and $S$ are the spin and symmetry factors, $E_{i,j}$ the energy of incoming particle, and $\mathbf{p}_f$ the 3-momentum of final-state particle.
Similarly, the thermally-averaged cross section $\langle \sigma v^2 \rangle$ for the $i,j,k\rightarrow l,m$ process is given by 
\begin{align}
\langle \sigma v^2 \rangle = \frac{1}{S}\frac{1}{F}\sum_{\rm spins}\frac{1}{128\pi^2 E_iE_jE_k(E_i+E_j+E_k)}\int d\Omega |\mathbf{p}_f||\mathcal{M}|^2\,.
\end{align}
For the decay rate, see Appendix \ref{apdx:XRdecay}.
We utilise \texttt{FeynRules} \cite{Alloul:2013bka} and \texttt{CalcHEP} \cite{Belyaev:2012qa} to compute the scattering amplitudes. We work in the nonrelativistic framework and take only $s$-wave contributions.

Instead of $n_i$ and $t$, it is convenient to use the yield $Y_i$ and the dimensionless time variable $x$, defined as
\begin{align}
Y_{i} \equiv n_{i}/s(T)
\,,\qquad
x \equiv m_i/T\,,
\end{align}
where $s(T)=(2\pi^2/45)g_{*s}T^3$ is the entropy density with $T$ being the temperature of the heat bath. Note that, during the radiation-dominated era,
\begin{align}
dt = \frac{x}{H(m_i)}dx\,,
\end{align}
where $H(m_i) = \sqrt{(\pi^2/90)g_*}m_i^2/M_{\rm P}$. Note also that $s(T) = s(m_i)/x^3$.
Denoting $s(m_i) \equiv s$ and $H(m_i) \equiv H$, the Boltzmann equation can be recast into
\begin{align}
\frac{dY_i}{dx} &=
-\frac{x}{H}\langle \Gamma \rangle_{i\rightarrow j,k}
\left(
Y_i - Y_i^{\rm eq}\frac{Y_j Y_k}{Y_j^{\rm eq} Y_k^{\rm eq}}
\right)
-\frac{x}{H}\langle \Gamma \rangle_{i\rightarrow j,k,l}
\left(
Y_i - Y_i^{\rm eq}\frac{Y_j Y_k Y_l}{Y_j^{\rm eq} Y_k^{\rm eq} Y_l^{\rm eq}}
\right)
\nonumber\\
&\quad
-\frac{s}{Hx^2}\langle \sigma v \rangle_{i,j \rightarrow l,m}
\left(
Y_i Y_j - Y_i^{\rm eq}Y_j^{\rm eq}\frac{Y_l Y_m}{Y_l^{\rm eq}Y_m^{\rm eq}}
\right)
\nonumber\\
&\quad
-\frac{s^2}{Hx^5}\langle \sigma v^2 \rangle_{i,j,k \rightarrow l,m} 
\left(
Y_i Y_j Y_k - Y_i^{\rm eq}Y_j^{\rm eq}Y_k^{\rm eq}
\frac{Y_l Y_m}{Y_l^{\rm eq} Y_m^{\rm eq}}
\right)
\,.
\end{align}
The full Boltzmann equations are summarised in Appendix \ref{apdx:BoltzmannEqs}. We choose $m_i = m_Y$.

The DM relic density is then given by
\begin{align}
\Omega_{\rm DM}h^2 = 
\Omega_y h^2 + \Omega_{X_I} h^2 + \Omega_{X_R} h^2
=\frac{s_0 h^2}{\rho_c}\left(
m_Y Y_y^0 + m_{X_I} Y_{X_I}^0 + m_{X_R} Y_{X_R}^0
\right)
\,,
\end{align}
where the subscript ``0'' indicates the present value, $\rho_c$ is the critical energy density, and $h\approx 0.67$ is the scaling factor for the Hubble parameter. Here $Y_y \equiv 2 Y_Y = 2Y_{Y^*}$.

The input parameters for the model under consideration are chosen as
\begin{gather*}
m_{X_R} \,,\quad
m_{X_I} \,,\quad
m_Y \,,\quad
m_{Z^\prime} \,,\quad
m_{h^\prime} \,,\nonumber\\
g_X \,,\quad
\lambda_Y \,,\quad
\lambda_X \,,\quad
\lambda_{XY} \,,\quad
\lambda_{Y\phi} \,,\quad
\lambda_{X\phi} \,,\quad
\lambda^\prime_{Y\phi} \,,\nonumber\\
\epsilon \,,\quad
\alpha\,.
\end{gather*}
In the following, we fix $m_{Z^\prime} = 200$ MeV, $m_{h^\prime} = 30$ GeV, $\lambda_X = 0.025$, and $\epsilon=2\times10^{-4}$. We assume that all the quartic and portal couplings are non-negative. Furthermore we assume that $\lambda^\prime_{Y\phi}$ is positive. As explained in the previous section, $\lambda_{XY}$ is required to be small in order to implement multi-component DM scenarios in the large-$m_{h^\prime}$ limit. We first perform the analysis with $\lambda_{XY} = 0$ and give a maximum value of $\lambda_{XY}$ that do not jeopardise the multi-component scenarios realised by our chosen benchmark points. We then present another sets of parameter values with a somewhat smaller $m_{h^\prime}$ together with large dark Higgs-portal couplings, $\lambda_{X\phi}$ and $\lambda_{Y\phi}$. In this case the destructive interference occurs and $\lambda_{XY}$ may take a larger value. The mixing angle between the SM Higgs and the dark Higgs is chosen to be small, $\alpha = 10^{-2}$, to evade the invisible Higgs decay constraint. Consequently, couplings to the SM Higgs field are not important.
Before presenting our results for the multi-component DM scenarios, let us consider theoretical and experimental constraints.

\subsection{Constraints}
\label{subsec:constraints}
\begin{figure}[tp]
\begin{center}
\includegraphics[scale=0.57]{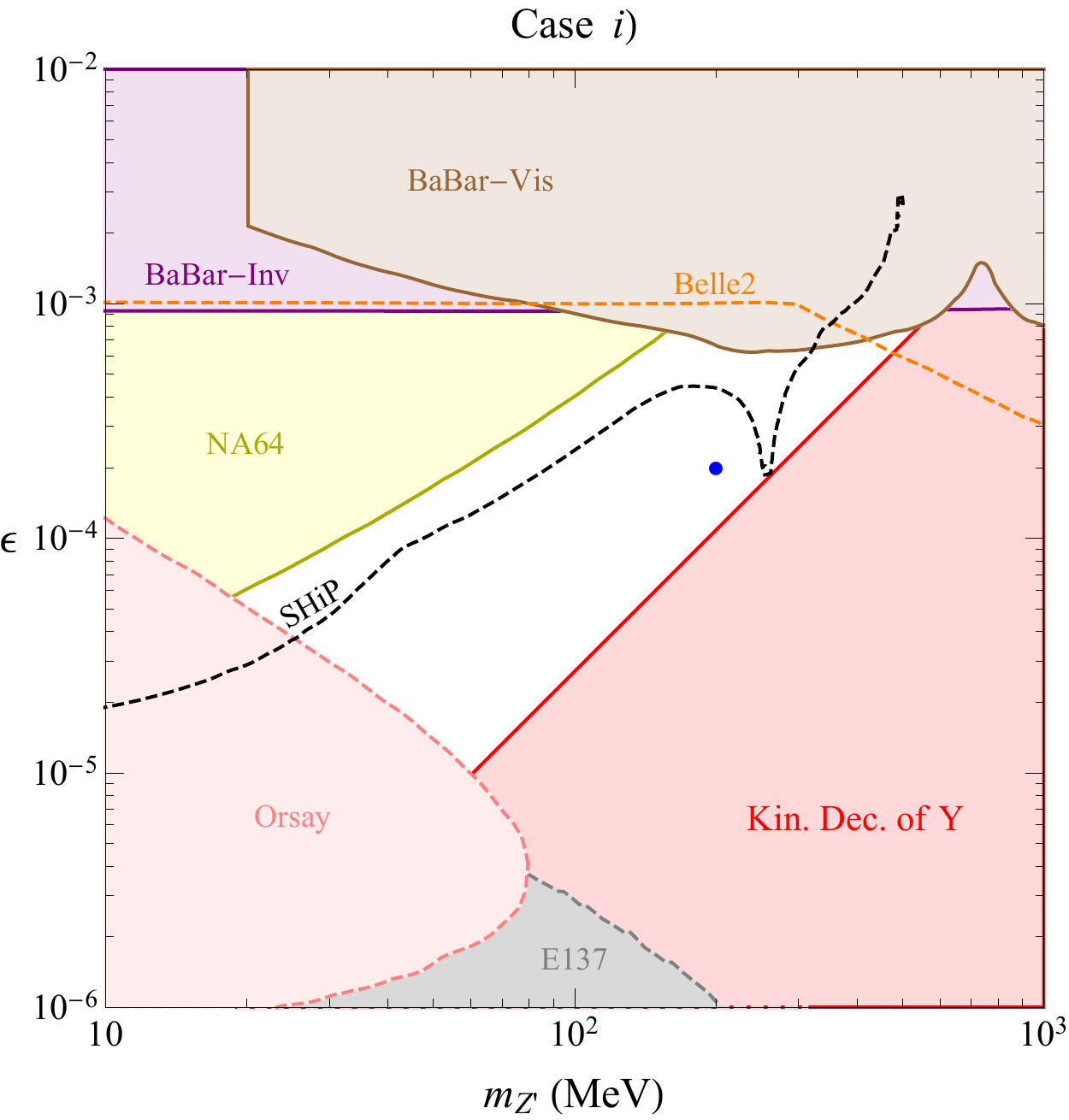}
\includegraphics[scale=0.57]{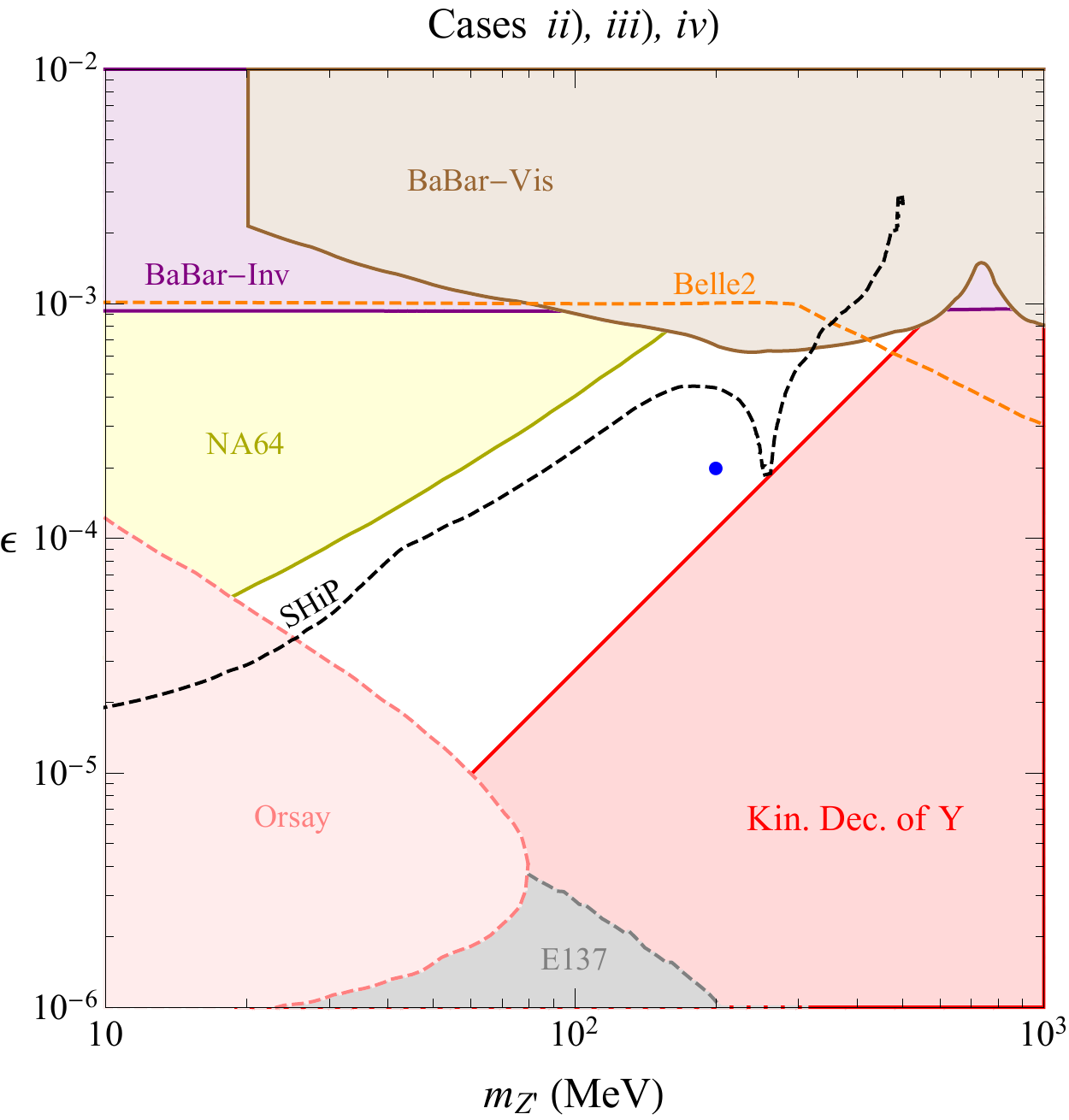}
\caption{Kinetic decoupling and $Z^\prime$-search constraints for the case {\it i)} (left) and cases {\it ii)}, {\it iii)}, and {\it iv)} (right) in the ($m_{Z^\prime}$, $\epsilon$) plane. See the main text in Section \ref{subsec:constraints} for details of the used constraints. We have the same plot for {\it ii)}, {\it iii)}, and {\it iv)} since the same $m_Y$, $m_{Z^\prime}$, and $\epsilon$ are used. Furthermore, we do not see a clearly visible difference between the left and right plots since the difference in $m_Y$ is small. The blue points represent our benchmark cases shown in Table \ref{tab:inputparams}.}
\label{fig:constraints}
\end{center}
\end{figure}

\begin{figure}[tp]
\begin{center}
\includegraphics[scale=0.57]{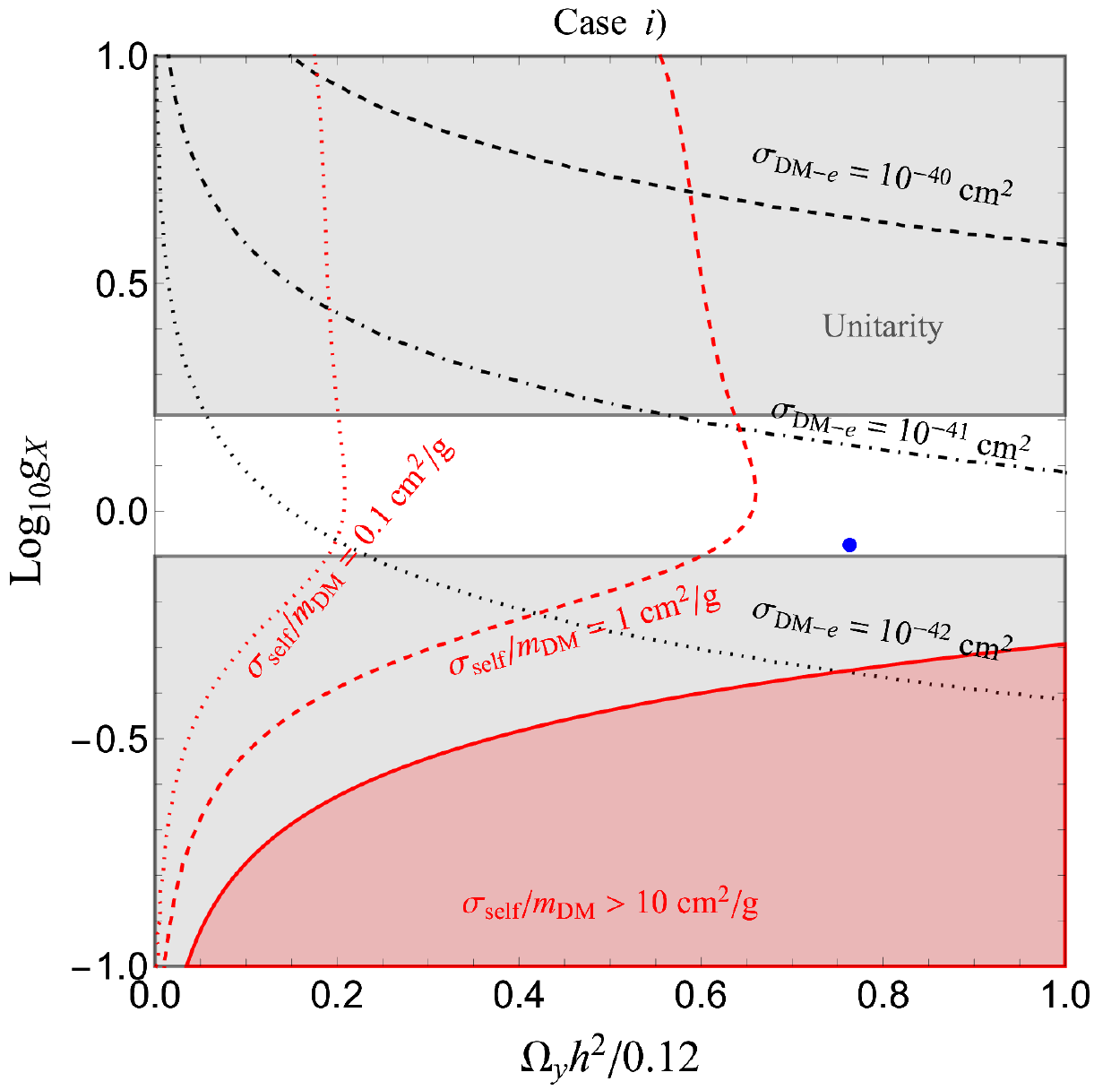}
\includegraphics[scale=0.57]{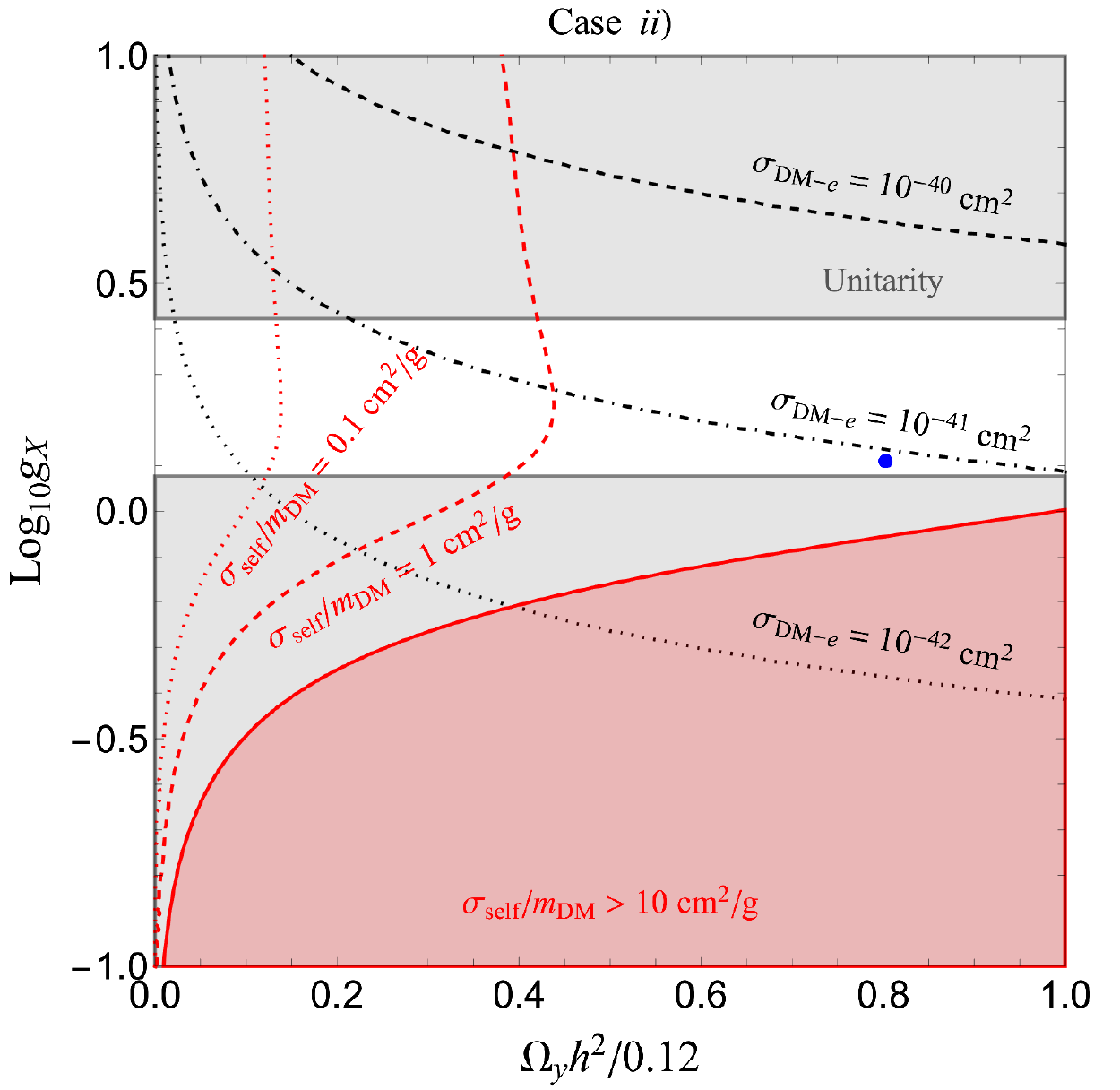}\\
\includegraphics[scale=0.57]{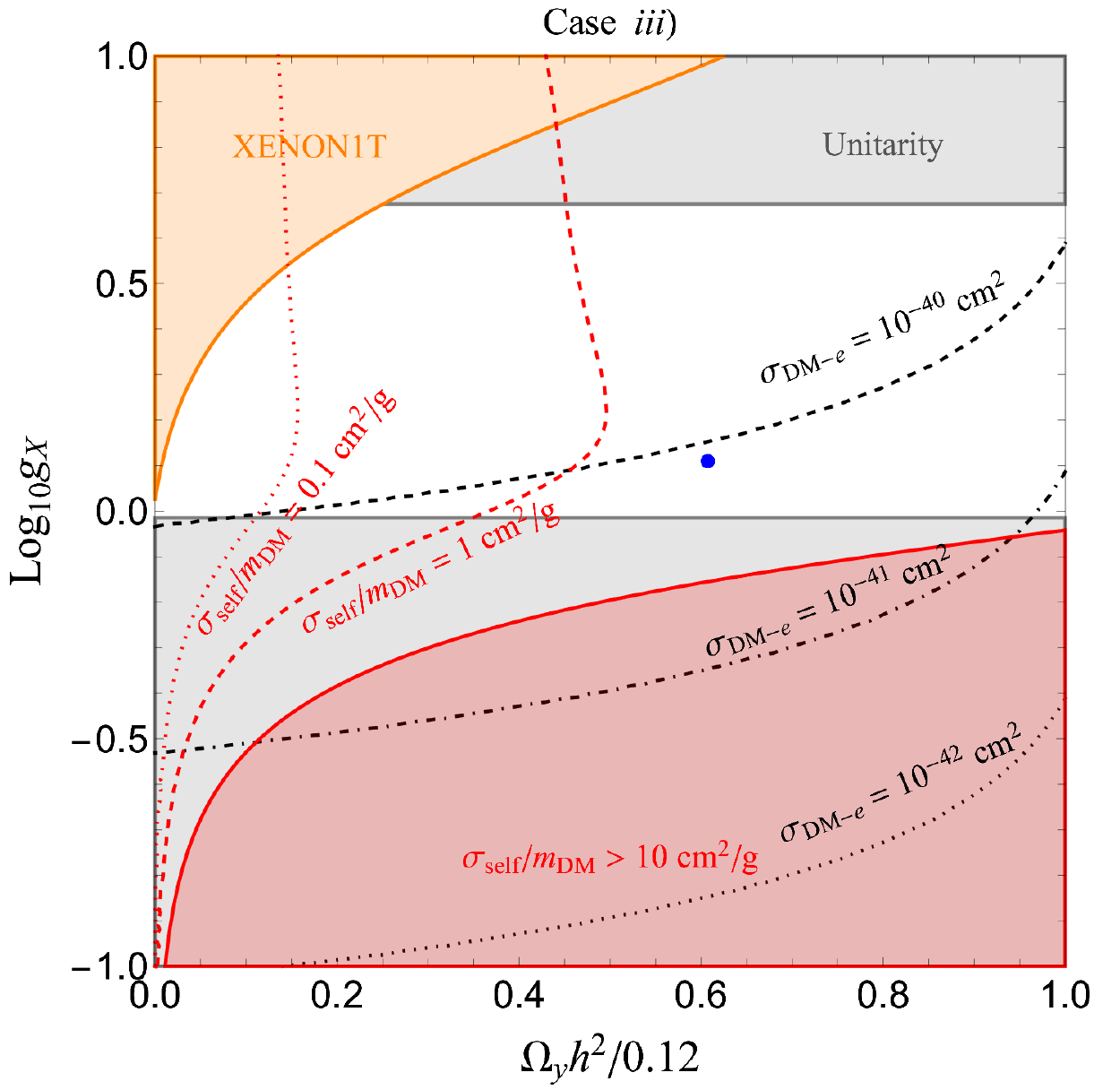}
\includegraphics[scale=0.57]{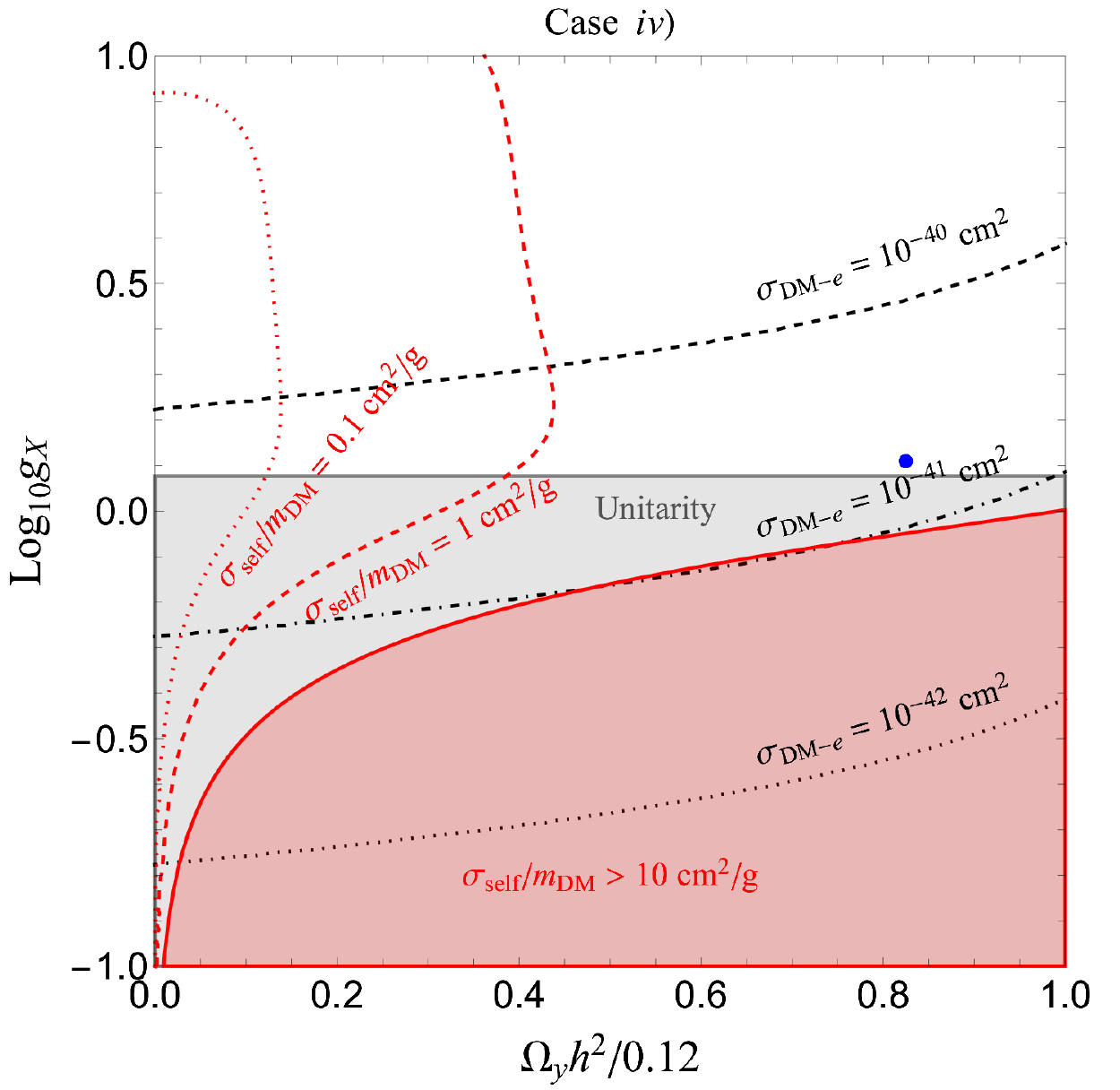}
\caption{For each case, the unitarity bound (grey), direct detection bound (orange), and large self-scattering cross section bound $\sigma_{\rm self}/m_{\rm DM} > 10$ ${\rm cm^2/g}$ (red) are shown for different values of the dark gauge coupling $g_X$ and the fraction of the $Y$ relic density $\Omega_y h^2/0.12$. The black lines represent DM--electron scattering cross section values, $10^{-40}$ ${\rm cm^2}$ (dashed), $10^{-41}$ ${\rm cm^2}$ (dot-dashed), and $10^{-42}$ ${\rm cm^2}$ (dotted) from top to bottom, respectively. The red lines correspond to two different values of the self-scattering cross section, $0.1$ ${\rm cm^2/g}$ (dotted) and $1$ ${\rm cm^2/g}$ (dashed). For details of the used constraints, see the main text in Section \ref{subsec:constraints}. The blue points represent our benchmark cases shown in Table \ref{tab:inputparams}. Here we vary only $g_X$ and the fraction of the $Y$ relic density, with the condition $(\Omega_y +\Omega_{X_I} + \Omega_{X_R}) h^2= 0.12$. For the two-component DM cases {\it i)} and {\it ii)}, $\Omega_{X_R} = 0$ is chosen, while for the three-component DM cases {\it iii)} and {\it iv)}, $\Omega_{X_R} = \Omega_{X_I}$ is assumed, except for the self-scattering cross section for which we follow the method outlined in Section \ref{subsec:constraints}. The other input parameters are fixed (see Table \ref{tab:inputparams}). We note that the correct present-day DM relic density constraint is not strictly imposed, except at the benchmark points. Thus, it is important to note that not all the white region is allowed. On the other hand, this approach allows one to comprehensively understand the various constraints and where our benchmark cases lie. The full picture requires an intensive parameter scan by numerically solving the coupled Boltzmann equations which is beyond the scope of the current work.}
\label{fig:constraints2}
\end{center}
\end{figure}

We take into account the following theoretical and experimental constraints. 
\begin{itemize}
\item Perturbativity\\
The perturbativity condition imposes the following conditions for the coupling parameters:
\begin{align}
\lambda_i < 4\pi \,,\quad
\lambda^\prime_{Y\phi} < 4\pi \,,\quad
g_X < 4\pi\,,
\end{align}
where $i=\{X,Y,XY,X\phi,Y\phi,XH,YH,\phi H\}$.

\item Unitarity\\
The unitarity bounds are given by
\begin{align}
|\mathcal{M}|_{i,j \rightarrow l,m} < 8\pi\,,
\end{align}
for $i,j,l,m=\{X_I,X_R,Y\}$. The unitarity bound is shown in Fig.~\ref{fig:constraints2} in the ($\Omega_y h^2, g_X$) plane.

\item Kinetic equilibrium\\
If the $3\rightarrow 2$ processes are dominant, the kinetic energies of the DM candidates keep piling up.
Thus, the kinetic energies need to be released. This can be achieved when the DM particles have a way to talk to the SM thermal bath, so that elastic scattering processes release the kinetic energy. See Refs.~\cite{Choi:2019zeb,Gondolo:2012vh,Kuflik:2015isi} for details.
Following Ref.~\cite{Gondolo:2012vh}, we compute the momentum relaxation rate $\gamma$ of a DM particle scattered off the relativistic SM thermal bath species, namely the electron. We then require the relaxation rate to be larger than or compatible with the kinetic energy production from the $3\rightarrow2$ processes, i.e. $\gamma \gtrsim H x_{\rm KD}^2$, until the kinetic decoupling time $x_{\rm KD} \equiv m_Y/T_{\rm KD}$. We set the kinetic decoupling time to be similar to the freeze-out time, $x_f$.
In our study, we use the $Z'$-portal interaction to communicate with the relativistic SM heat bath. Since the relativistic SM degree of freedom is electron at $T\sim O({\rm MeV})$, we consider the DM--$e$ elastic scattering process.
The most dominant $3\rightarrow 2$ processes are the $Y$-involving processes, $Y,Y,Y\rightarrow Y,Y^*$ and $Y,Y^*,Y^* \rightarrow Y,Y$. Thus we focus on the kinetic energy release of the field $Y$. From the kinetic equilibrium condition we see that the $Z'$ mass is bounded as follows:
\begin{align}
m_{Z'} \lesssim \left(
\frac{31\pi^3 e^2 c_w^2 g_X^2}{1701x_{\rm KD}^6}
\right)^{1/4}\left(
m_Y^3 M_{\rm P}
\right)^{1/4} \sqrt{\epsilon}
\,.
\end{align}
The kinetic decoupling region is sketched in Fig.~\ref{fig:constraints}.

\item DM relic density\\
The current DM relic density is given by \cite{Aghanim:2018eyx}
\begin{align}
\Omega_{\rm DM}h^2 = 0.1200 \pm 0.0012\,.
\end{align}

\item Invisible Higgs decay\\
The Higgs invisible decay is bounded \cite{Sirunyan:2018owy} as
\begin{align}
{\rm BR}_{h} < 0.19\,,
\end{align}
at 90\% C.L., where ${\rm BR}_h$ is branching ratio of the Higgs invisible decay.
We choose a small enough mixing angle $\alpha=10^{-2}$ to satisfy the bound.

\item Direct detection\\
For the DM direct detection bounds, we consider Xenon10 \cite{Essig:2017kqs}, Xenon1T \cite{Aprile:2019jmx,Aprile:2019xxb}, and expected SENSEI-100 bound \cite{Battaglieri:2017aum}. 
For all the benchmark scenarios that we consider (Table \ref{tab:inputparams}), the DM-electron cross sections are tiny, and thus safe from the direct detection constraints. The DM direct detection bounds are depicted in Fig.~\ref{fig:constraints2}.
The DM-quark scattering bounds are unimportant since the DM masses are less than 1 GeV in our benchmark scenarios.

\item $Z^\prime$ searches\\
We use the BaBar visible/invisible decay of $Z^\prime$ constraints \cite{Lees:2014xha,Lees:2017lec}, Belle2 invisible decay of $Z^\prime$ constraint \cite{Essig:2013vha}, and beam dump constraints coming from NA64 \cite{Banerjee:2019hmi,Gninenko:2019qiv}, SHiP \cite{SHiP:2020noy}, E137 \cite{Bjorken:1988as}, and Orsay \cite{Davier:1989wz}; see also Refs.~\cite{Andreas:2012mt,Marsicano:2018krp}. These constraints impose bounds on the kinetic mixing angle $\epsilon$, depending on the $Z^\prime$ mass, and are depicted in Fig.~\ref{fig:constraints}.

\item DM self-scattering cross section\\
The SIMP-type DM models naturally have a sizeable self-scattering cross section that helps to solve the small-scale problems mentioned in Section \ref{sec:intro}. We require \cite{Spergel:1999mh,Elbert:2014bma,Tulin:2017ara,Chu:2018fzy}
\begin{align}
0.1 \; {\rm cm^2/g} < \frac{\sigma_{\rm self}}{m_{\rm DM}} < 10 \; {\rm cm^2/g}\,.
\end{align}
The Bullet Cluster \cite{Markevitch:2003at,Clowe:2003tk} imposes a stricter constraint, $\sigma_{\rm self}/m_{\rm DM} < 1\;{\rm cm^2/g}$. See also Ref. \cite{Randall:2007ph}.
A similar bound is obtained from cosmological simulations with self-interacting DM on the scales of galaxies and galaxy clusters \cite{Rocha:2012jg,Peter:2012jh}.

For the $m_{\rm DM}$, we take an effective DM mass. For the cases {\it i)} and {\it iii)} the dominant contribution to $\sigma_{\rm self}$ comes from the $Y$ self-scattering cross section, as $X_I$ and $X_R$ are much heavier than $Y$. We thus choose $m_{\rm DM} = m_Y$. For the case of {\it ii)} both $X_I$ and $Y$ may contribute; thus we choose $m_{\rm DM} = (m_Y + m_{X_I})/2$. The case {\it iv)} is more complicated. In this case $X_R$, $X_I$, and $Y$ all contribute. We thus consider all the contributions with the effective masses. All the contributions are weighted by the fractions of the relic density. In summary,
\begin{align}
&\frac{\sigma_{\rm self}}{m_{\rm DM}}
\bigg\vert_{{\it i)},{\it iii)}} &&\hspace{-2mm}=\;\;
\frac{1}{4}\left(
\frac{\Omega_y}{\Omega_{\rm DM}}
\right)^2 \frac{1}{m_Y} \left(
2\sigma_{Y,Y\rightarrow Y,Y}
+\sigma_{Y,Y^* \rightarrow Y,Y^*}
\right)
\,,\\
&\frac{\sigma_{\rm self}}{m_{\rm DM}}
\bigg\vert_{{\it ii)}} &&\hspace{-2mm}=\;\;
\frac{\sigma_{\rm self}}{m_{\rm DM}}\bigg\vert_{{\it i)},{\it iii)}}
+\left(
\frac{\Omega_{X_I}\Omega_{y}}{\Omega_{\rm DM}^2}
\right)
\frac{2}{m_Y+m_{X_I}}
\sigma_{X_I,Y \rightarrow X_I,Y}
\nonumber\\
&&&\quad
+\left(
\frac{\Omega_{X_I}}{\Omega_{\rm DM}}
\right)^2
\frac{1}{m_{X_I}}
\left(
\sigma_{X_I,X_I\rightarrow X_I,X_I}
+\sigma_{X_I,X_I\rightarrow Y,Y^*}
\right)
\,,\\
&\frac{\sigma_{\rm self}}{m_{\rm DM}}
\bigg\vert_{{\it iv)}} &&\hspace{-2mm}=\;\;
\frac{\sigma_{\rm self}}{m_{\rm DM}}\bigg\vert_{{\it ii)}}
+\left(
\frac{\Omega_{X_R}\Omega_{Y}}{\Omega_{\rm DM}^2}
\right)
\frac{2}{m_{X_R}+m_{Y}}
\left(
\sigma_{X_R,Y \rightarrow X_R,Y}
+\sigma_{X_R,Y \rightarrow X_I,Y}
\right)
\nonumber\\
&&&\quad
+\left(
\frac{\Omega_{X_R}\Omega_{X_I}}{\Omega_{\rm DM}^2}
\right)
\frac{2}{m_{X_R}+m_{X_I}}
\left(
\sigma_{X_I,X_R \rightarrow X_I,X_R}
+\sigma_{X_I,X_R \rightarrow Y,Y^*}
\right)
\nonumber\\
&&&\quad
+\left(
\frac{\Omega_{X_R}}{\Omega_{\rm DM}}
\right)^2
\frac{1}{m_{X_R}}
\left(
\sigma_{X_R,X_R\rightarrow X_R,X_R}
+\sigma_{X_R,X_R\rightarrow X_I,X_I}
+\sigma_{X_R,X_R\rightarrow Y,Y^*}
\right)
\,.
\end{align}
In Fig.~\ref{fig:constraints2}, $\sigma_{\rm self}/m_{\rm DM} = 0.1$ ${\rm cm^2/g}$, $1$ ${\rm cm^2/g}$, and $\sigma_{\rm self}/m_{\rm DM}>10$ ${\rm cm^2/g}$ are shown.
\end{itemize}

Taking all of the listed constraints into account we now present four benchmark results for the multi-component DM scenarios.


\subsection{Two-component Scenario}
\label{subsec:2compnt}
When the mass of $X_R$ is sufficiently larger than $m_{X_I} + m_{Z^\prime}$, $X_R$ cannot be a DM candidate as it decays into $X_I$ and $Z^\prime$. Thus we have the two-component scenario.
We consider two cases: {\it i)} $m_{X_R} \gg m_{X_I} \gg m_Y$ and {\it ii)} $m_{X_R} \gg m_{X_I} \approx m_Y$. The input parameter values are shown in Table \ref{tab:inputparams}.

\begin{table}[tp]
\begin{center}
\begin{tabular}{|c|c|c|c|c|c|c|c|c|}
\hline
Case & $m_{X_R}$ [MeV] & $m_{X_I}$ [MeV] & $m_Y$ [MeV] &
$g_X$ & $\lambda_Y$ & $\lambda_{Y\phi}$ & 
$\lambda_{X\phi}$ & $\lambda^\prime_{Y\phi}$
\\\hline
{\it i)} & 800 & 200 & 50 &
0.85 & 6.41 & 0.9 & 0.9 & 0.255 \\
\hline
{\it ii)} & 400 & 40 & 37.5 &
1.3 & 6.27 & 1.3 & 0.5 & 0.295 \\
\hline
{\it iii)} & 150.01 & 150 & 37.5 &
1.3 & 5.54 & 0.05 & 0.05 & 0.271 \\
\hline
{\it iv)} & 40.001 & 40 & 37.5 &
1.3 & 6.27 & 2.65 & 2.2 & 0.295 \\
\hline
\end{tabular}
\end{center}
\caption{Input parameter values for the four benchmark cases. Cases {\it i)} and {\it ii)} correspond to the two-component scenarios and cases {\it iii)} and {\it iv)} correspond to the three-component scenarios. We chose $m_{Z^\prime}=200$ MeV, $m_{h^\prime}=30$ GeV, $\lambda_X = 0.025$, $\alpha = 10^{-2}$, and $\epsilon=2\times10^{-4}$. Our four benchmark cases are marked as blue points in Figs.~\ref{fig:constraints} and \ref{fig:constraints2}.}
\label{tab:inputparams}
\end{table}

We solve the Boltzmann equations given in Appendix \ref{apdx:BoltzmannEqs} for the given parameters, and the solutions are shown in Fig.~\ref{fig:BEsols}. We see that, in the case of $m_{X_I} \gg m_Y$, $X_I$ freezes out first at $x \approx 5$, followed by the $Y$ freeze-out at $x \approx 20$. When $m_{X_I} \approx m_Y$, both $X_I$ and $Y$ freeze out at almost the same time $x \approx 20$. In the case {\it i)}, we find that $X_I$ ($Y$) constitutes about 23.7\% (76.3\%) of the total relic density, while in the case {\it ii)}, the fractions of $X_I$ and $Y$ relics are respectively 19.7\% and 80.3\%. Thus we identify these scenarios as two-component DM scenarios. The self-scattering cross sections are given by $1.49$ ${\rm cm^2/g}$ and $3.83$ ${\rm cm^2/g}$ for cases ${\it i)}$ and ${\it ii)}$, respectively.
We summarise the results in Table \ref{tab:results}.

\subsection{Three-component Scenario}
\label{subsec:3compnt}
When the mass gap between $X_R$ and $X_I$ is small enough, $X_R$ may become a good DM candidate.
We take the mass splitting to be small enough, $m_{X_R} - m_{X_I} \ll {\rm min}\{2m_Y,m_{Z^\prime}\}$,  such that the $X_R$ decay channels such as $X_R \rightarrow X_I,Z^\prime$ and $X_R \rightarrow X_I,Y,Y^*$ are kinematically closed \footnote{
Note that the decay channel $X_R \rightarrow X_I, \ell, \bar{\ell}$, where $\ell$ is the SM fermion, can still remain open through a virtual $Z/Z^\prime$ exchange. For the chosen mass gap of $\lesssim 10$ keV, $\ell=\nu$ is possible. However, for our benchmark cases with such a small gap, the lifetime of $X_R$,
\begin{align*}
\Gamma_{X_R\rightarrow X_I,\nu,\bar{\nu}} \approx
6.06\times 10^{-46} \;{\rm GeV} \;
\left(
\frac{\epsilon}{2\times 10^{-4}}
\right)^2 g_X^2 \left(
\frac{m_{X_R}-m_{X_I}}{10\;{\rm keV}}
\right)^5\,,
\end{align*}
is longer than the age of the Universe as pointed out in Ref. \cite{Baek:2020owl}. We shall thus ignore this decay channel in the following.
Note also that the partial width of the $X_R$ decaying into $X_I$ and three photons, as shown in Ref. \cite{Harigaya:2020ckz}, is around twelve orders of magnitude smaller than $\Gamma_{X_R\rightarrow X_I,\nu,\bar{\nu}}$.
}. 
Thus, together with $Y$ and $X_I$, we have a three-component scenario.
We consider two cases: {\it iii)} $m_{X_R} \approx m_{X_I} \gg m_Y$ and {\it iv)} $m_{X_R} \approx m_{X_I} \approx m_Y$. The input parameter values are shown in Table \ref{tab:inputparams}.
We chose a large dark gauge coupling, $g_X\sim\mathcal{O}(1)$.

While the decay channels of the field $X_R$ are kinematically forbidden, it is still possible that the field $X_R$ may be converted into the field $X_I$ via the conversion process $X_R,Y\rightarrow X_I,Y$. We numerically solve the full Boltzmann equation and found that the mass gap should be less than $\sim$10 keV, i.e. $m_{X_R} - m_{X_I} \lesssim 10$ keV to suppress such a process, thereby achieving three-component scenarios.

\begin{figure}[tp]
\begin{center}
\includegraphics[scale=0.55]{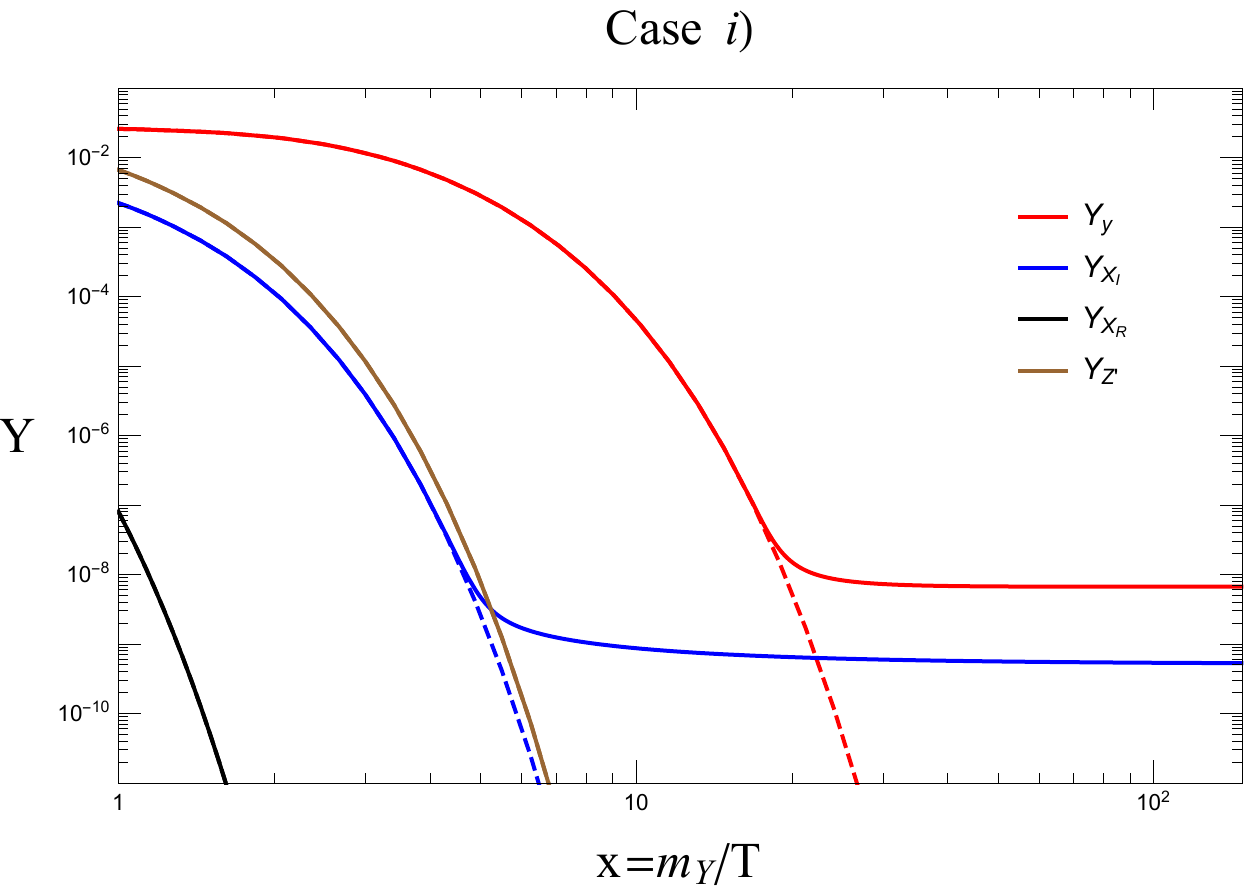}\;
\includegraphics[scale=0.55]{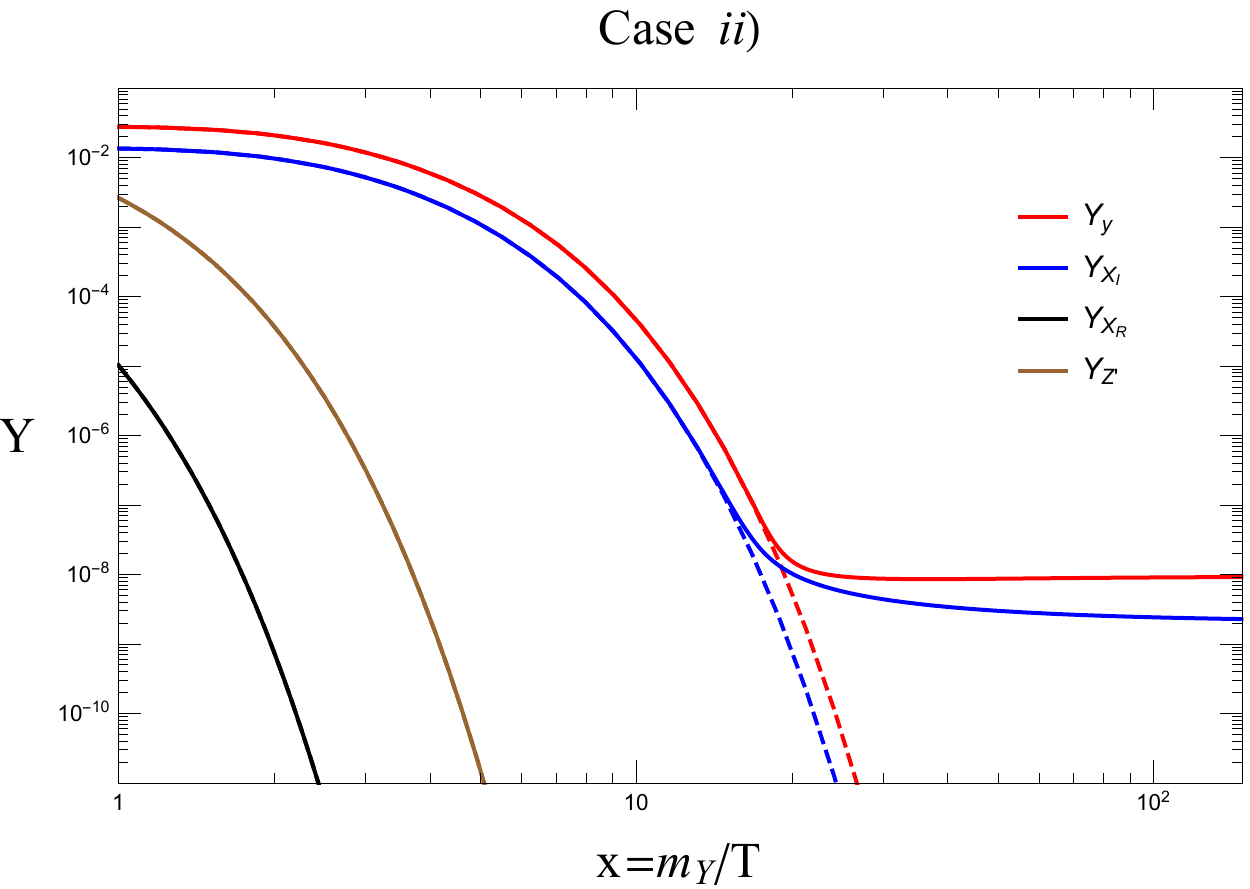}\\
\includegraphics[scale=0.55]{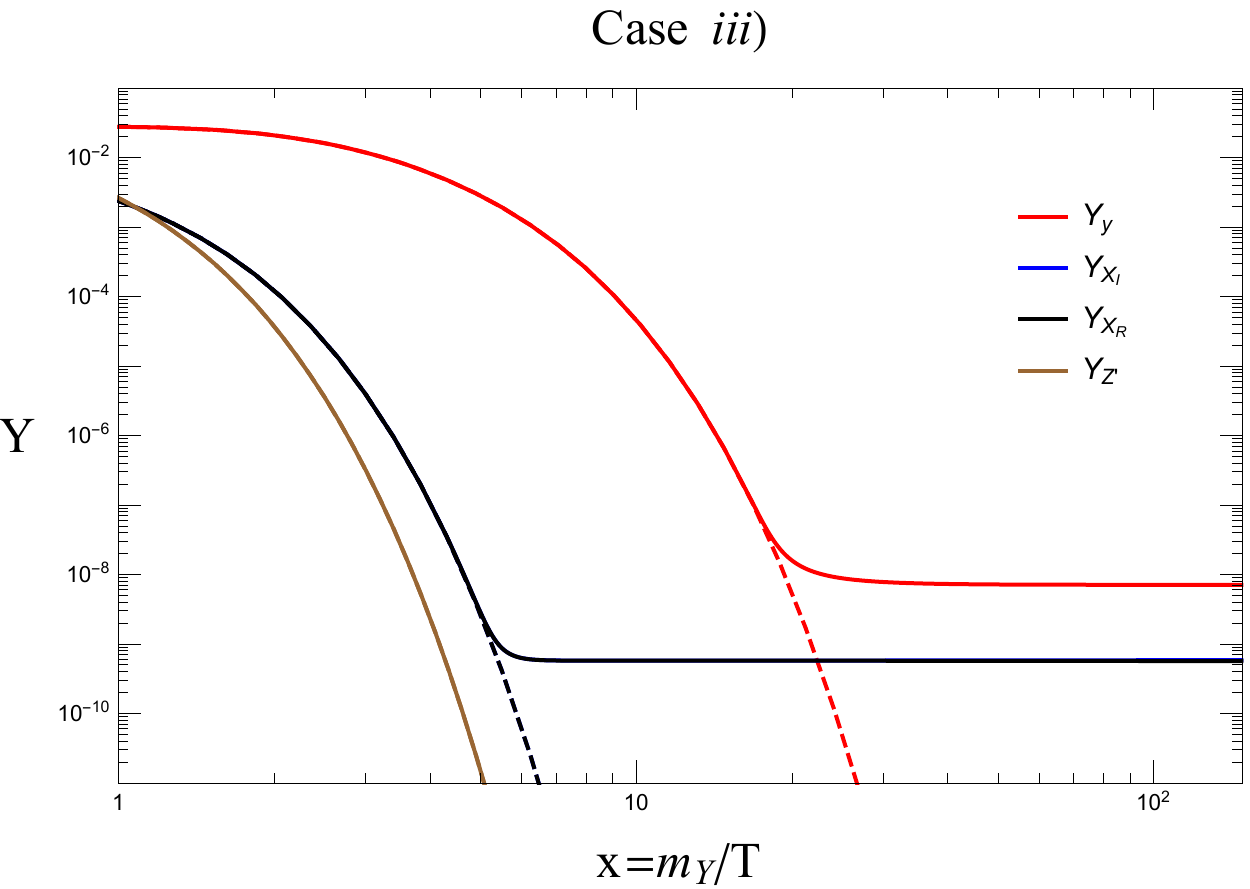}\;
\includegraphics[scale=0.55]{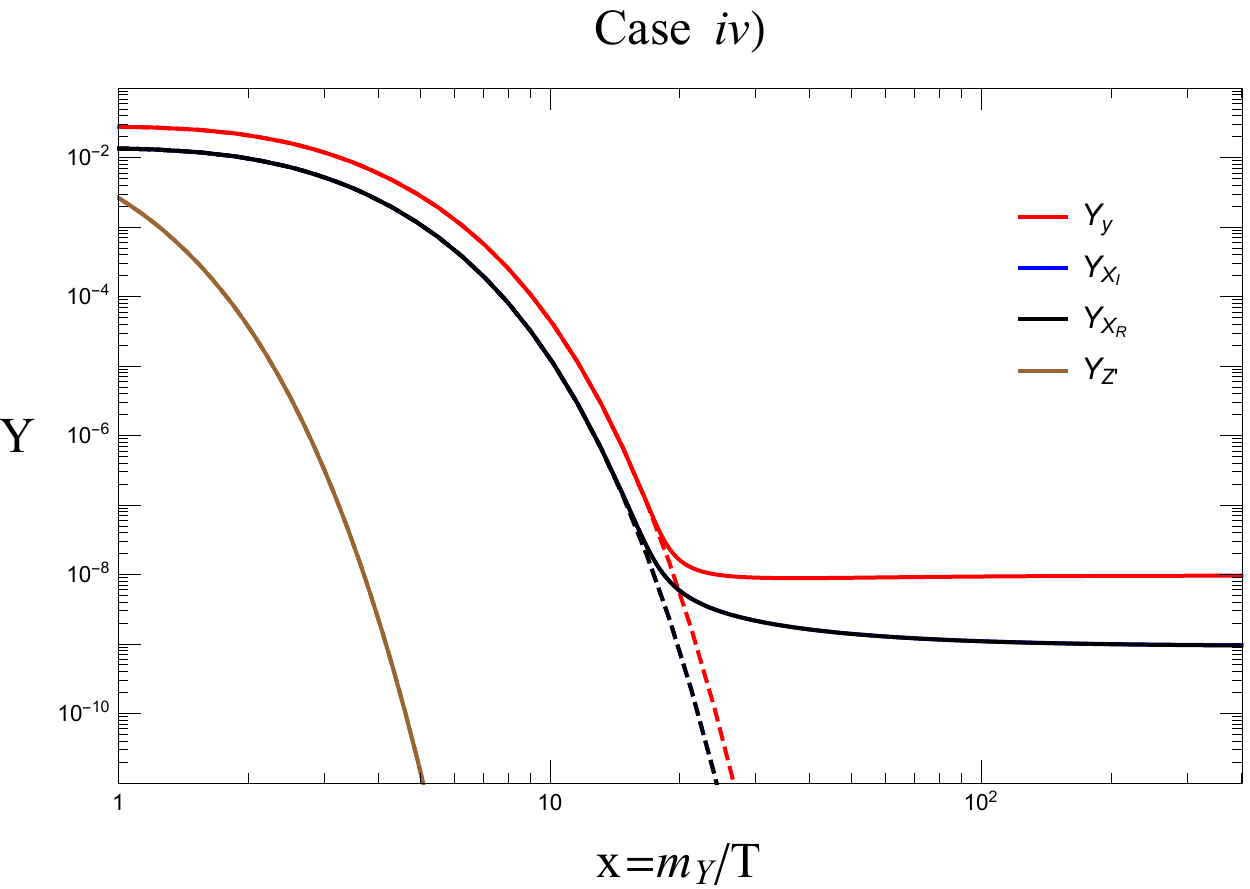}
\caption{The solutions of the Boltzmann equations summarised in Appendix \ref{apdx:BoltzmannEqs} are shown for cases {\it i)}--{\it iv)}. The red, blue, black, and brown solid lines are the yields of $Y$, $X_I$, $X_R$, and $Z^\prime$, respectively; here $Y_y \equiv 2Y_Y = 2Y_{Y^*}$. The dashed lines correspond to the equilibrium states. In the cases {\it i)} and {\it ii)} $Y$ and $X_I$ relics become frozen out, indicating two-component DM scenarios. On the other hand, in the cases {\it iii)} and {\it iv)} all the DM candidates freeze out, and thus we have three-component scenarios. We note that there is no visible difference between the $X_I$ relic and $X_R$ relic in the cases {\it iii)} and {\it iv)} due to the small mass gap. A large gauge coupling $g_X \sim \mathcal{O}(1)$ is chosen for our benchmark points. In this case, once the mass gap becomes larger than $\sim$10 keV, a significant amount of the relic of $X_R$ is converted into the $X_I$ relic, mainly through $X_R,Y \rightarrow X_I,Y$, becoming a two-component scenario. In all cases we see that $Z^\prime$ follows its equilibrium state.}
\label{fig:BEsols}
\end{center}
\end{figure}

The solutions of the Boltzmann equations are shown in Fig.~\ref{fig:BEsols}.
Since $m_{X_R} \approx m_{X_I}$ we do not see any clear deviations between the $X_R$ evolution and $X_I$ evolution. Similar to the cases {\it i)} and {\it ii)}, we see that, in the case of $m_{X_I} \gg m_Y$, $X_I$ freezes out first at $x \approx 5$, followed by the $Y$ freeze-out at $x \approx 20$. When $m_{X_I} \approx m_Y$, both $X_I$ and $Y$ freeze out at almost the same time $x \approx 20$.
In the case {\it iii)}, we find that $X_I$ and $X_R$ ($Y$) constitute about 20\% each (60\%) of the total relic density, while in the case {\it iv)}, the fractions of $X_I$ and $X_R$ relics are about 9\% each, with 82\% of $Y$ relic. Thus we identify these scenarios as three-component DM scenarios. The self-scattering cross sections are given by $1.66$ ${\rm cm^2/g}$ and $4.04$ ${\rm cm^2/g}$ for cases ${\it iii)}$ and ${\it iv)}$, respectively.
We summarise the results in Table \ref{tab:results}.

\begin{table}[tp]
\begin{center}
\begin{tabular}{|c|c|c|c|c|c|c|}
\hline
Case & $\Omega_{X_R}/\Omega_{\rm DM}$ & $\Omega_{X_I}/\Omega_{\rm DM}$ & $\Omega_y/\Omega_{\rm DM}$ & $\sigma_{\rm self}/m_{\rm DM}$ (${\rm cm^2/g}$) & DM scenario
\\\hline
{\it i)} & 0 & 0.237 & 0.763 & 1.49 & Two-component \\
\hline
{\it ii)} & 0 & 0.197 & 0.803 & 3.83 & Two-component \\
\hline
{\it iii)} & 0.186 & 0.207 & 0.607 & 1.66 & Three-component \\
\hline
{\it iv)} & 0.087 & 0.088 & 0.825 & 4.04 & Three-component \\
\hline
\end{tabular}
\end{center}
\caption{The fractions of relic density for each DM candidate field and the self-scattering cross sections are shown for the four benchmark cases. The input parameters are summarised in Table \ref{tab:inputparams}. The total DM relic density is $\Omega_{\rm DM}h^2 = 0.12$. As expected from the mass gap between $X_I$ and $X_R$, cases {\it i)} and {\it ii)} give rise to two-component DM scenarios, while three-component DM scenarios are realised in the cases {\it iii)} and {\it iv)}. The self-scattering cross sections are somewhat larger than 1 ${\rm cm^2/g}$ imposed by the Bullet Cluster constraint \cite{Markevitch:2003at,Clowe:2003tk,Randall:2007ph} (see also Refs. \cite{Rocha:2012jg,Peter:2012jh} where the similar bound is obtained from cosmological simulations with self-interacting DM), but well within the bound, 10 ${\rm cm^2/g}$.}
\label{tab:results}
\end{table}

\vspace{5mm}

Let us comment on the parameter $\lambda_{XY}$ before we conclude.
In our benchmark cases {\it i)}--{\it iv)} we chose $\lambda_{XY}$ to be zero so that the contact interaction between $X$ and $Y$ gets suppressed, helping realizations of the multi-component DM scenarios in the large-$m_{h^\prime}$ limit.
For the chosen parameter values of the benchmark cases, we present the upper bound on $\lambda_{XY}$ to achieve such multi-component scenarios.
We first require that $\Omega_Y/\Omega_{\rm DM} \lesssim 90\%$ be the condition for a multi-component DM scenario.
For the case {\it i)}, we obtained the maximum value of $\lambda_{XY} = 1.5\times 10^{-4}$ with $\lambda_Y = 5.45$, in which case the relic of the $Y$ ($X_I$) field consists 90\% (10\%) of the total relic density.
For the case {\it ii)}, the maximum value is given by $\lambda_{XY} = 4.7 \times 10^{-5}$ with $\lambda_Y = 4.62$. The $Y$ ($X_I$) relic consists 90\% (10\%) of the total relic density.
For the case {\it iii)} (case {\it iv)}), $\lambda_{XY} = 4.0 \times 10^{-4}$ ($3.8 \times 10^{-4}$) is the maximum value, with $\lambda_Y = 3.71$ (4.95) that gives $\Omega_Y/\Omega_{\rm DM} = 90\%$ and $\Omega_{X_{I,R}}/\Omega_{\rm DM} = 5\%$. Note that the value of $\lambda_Y$ is modified accordingly in order to satisfy the correct relic density of $\Omega_{\rm DM}h^2 = 0.12$. The rest of the input parameters are kept to be the same as in Table \ref{tab:inputparams}.
As such, the value of $\lambda_{XY}$ needs to be small to realise multi-component DM scenarios.
This is not a general statement, however, since the dark Higgs field is assumed to be massive compared to the other dark fields.
If the dark Higgs is chosen to be somewhat lighter and/or the couplings $\lambda_{X\phi}$ and $\lambda_{Y\phi}$ are somewhat larger, the value of $\lambda_{XY}$ may become larger, due to different signs in terms between the contact interaction and the dark Higgs mediated processes, causing cancellation, i.e. destructive interferences.
In Table \ref{tab:inputparamsprime} we present such cases; we choose $m_Y = 37.5$ MeV, $m_{Z^\prime} = 200$ MeV, $m_{h^\prime} = 10$ GeV, $g_X = 1$, $\lambda_{X} = 1$, $\lambda_{X\phi}=\lambda_{Y\phi} = 10$, and $\lambda_Y = 5$ for all the four cases.

\begin{table}[tp]
\begin{center}
\begin{tabular}{|c|c|c|c|c|c|c|c|c|}
\hline
Case & $\frac{m_{X_R}}{m_{X_I}}$ & $\frac{m_{X_I}}{m_Y}$ & 
$\lambda^\prime_{Y\phi}$ &
$\lambda_{XY}$ &
$\frac{\Omega_{X_R}}{\Omega_{\rm DM}}$ & 
$\frac{\Omega_{X_I}}{\Omega_{\rm DM}}$  & 
$\frac{\Omega_y}{\Omega_{\rm DM}}$ & 
$\frac{\sigma_{\rm self}}{m_{\rm DM}}$ (${\rm cm^2/g}$)
\\\hline
{\it i$'$)} & 4 & 4 & 
0.2 &
0.02 & 
0 & 
0.30 & 
0.70 & 
1.79 \\
\hline
{\it ii$'$)} & 10 & 1.07 & 
0.29 &
0.03 & 
0 & 
0.21 & 
0.79 & 
5.46 \\
\hline
{\it iii$'$)} & 
1.0001 & 4 & 
0.29 &
0.04 & 
0.24 & 0.28 & 0.48 &
2.07\\
\hline
{\it iv$'$)} & 
1.0001 & 
1.07  & 
0.29 &
0.04 & 
0.09 & 0.10 & 0.81 & 
5.74 \\
\hline
\end{tabular}
\end{center}
\caption{
Input parameters with non-zero $\lambda_{XY}$ and outcomes of DM relics and the self-scattering cross sections. The rest of the input parameters are chosen as follows: $m_Y = 37.5$ MeV, $m_{Z^\prime} = 200$ MeV, $m_{h^\prime} = 10$ GeV, $g_X = 1$, $\lambda_X = 0.1$, $\lambda_{X\phi} = \lambda_{Y\phi} = 10$, and $\lambda_Y = 5$. All the cases are free from the constraints listed in Section \ref{subsec:constraints}. We see that the multi-component DM scenarios are still realised with sizeable $\lambda_{XY}$ values. This is due to the destructive interference between the $X$--$Y$ contact interaction and the dark Higgs mediated processes.
}
\label{tab:inputparamsprime}
\end{table}

\section{Conclusions}
\label{sec:conclusions}

In this paper, we constructed multi-component dark matter scenarios with $U(1)_X$ dark gauge symmetry broken into $Z_2 \times Z_3$. Two independent types of dark matter fields, namely $X$ and $Y$, distinguished by their different charges under $Z_2 \times Z_3$, emerge as separately stable dark matter candidates. The relic densities of different dark matter fields are governed by interactions between the dark fields within the dark sector through number-changing $2\rightarrow 2$ and $3\rightarrow 2$ processes. 
Unlike many other SIMP models in the literature, $X$ and $Y$ are completely independent of each other, with different masses and spins in principle. 
Due to this reason we call our model a genuine multi-component SIMP dark matter scenario. 

A nonzero vacuum expectation value of the dark Higgs field gives mass splitting to the real and imaginary parts of the $X$ field.
When the mass gap between the real part $X_R$ and imaginary part $X_I$ is large, only the lighter field $X_I$ becomes a stable dark matter candidate which, together with $Y$, gives rise to a two-component dark matter scenario.
On the other hand, once the mass gap is small enough, both $X_R$ and $X_I$ may contribute to the dark matter relic density; we thus have a three-component scenario.
In our analysis, the dark gauge coupling $g_X$ is chosen to be $\mathcal{O}(1)$, and we found that the mass gap should be less than $\sim$10 keV, i.e. $m_{X_R} - m_{X_I} \lesssim 10$ keV, in order to achieve three-component scenarios.

The unbroken $Z_3$ symmetry ensures the stability of the $Y$ field, which becomes another dark matter candidate of the SIMP-type. Due to the interactions between $X_{I,R}$ and $Y$, both $X_I$ and $X_R$ may contribute to $Y$'s $3\rightarrow 2$ annihilation processes through the $Z^\prime$ and the dark Higgs. These kinds of processes open a new class of SIMP models.

We presented four choices of the parameter set for the multi-component dark matter scenarios, two for a two-component dark matter scenario, {\it i)} $m_{X_R} \gg m_{X_I} \gg m_Y$ and {\it ii)} $m_{X_R} \gg m_{X_I} \approx m_Y$, and two for a three-component dark matter scenario, {\it iii)} $m_{X_R} \approx m_{X_I} \gg m_Y$ and {\it iv)} $m_{X_R} \approx m_{X_I} \approx m_Y$, by solving the full coupled Boltzmann equations and considering both theoretical and experimental constraints. Figures~\ref{fig:constraints} and \ref{fig:constraints2} show a set of constraints.
The parameter choices are given in Table~\ref{tab:inputparams} and the solutions of the Boltzmann equations are shown in Fig.~\ref{fig:BEsols}. For each case we computed fractions of the relic density. The results, together with values of the self-scattering cross section, are summarised in Table~\ref{tab:results}. We found that, in the large-$m_{h^\prime}$ limit, the $\lambda_{XY}$ coupling, which dilutes the relic density of the $X$ field via the $X$--$Y$ contact interaction, needs to take a small value to realise a multi-component dark matter scenario.
On the other hand, the $\lambda_{XY}$ coupling may take a larger value once the dark Higgs field becomes lighter and the dark Higgs portal couplings, $\lambda_{X\phi}$ and $\lambda_{Y\phi}$, become larger. In this case destructive interferences occur and the dilution of the $X$ relic is suppressed even with a larger value of $\lambda_{XY}$. In Table \ref{tab:inputparamsprime} we present four parameter sets of this case.

Our findings of the multi-component SIMP-type dark matter scenarios are new and interesting. One of the most distinctive features of our model is that the different dark matter fields $X$ and $Y$ may have very different mass scales, while significantly contributing to the total dark matter relic density today, as demonstrated in the cases {\it i)} and {\it iii)}.
In the current work a scalar $Z_2$-charged field is considered. It is certainly possible to have a fermionic $Z_2$-charged field~\footnote{
Note that the $Z_3$-charged dark matter cannot be fermionic.
} as studied e.g. in Refs.~\cite{Baek:2014poa,Ko:2019wxq,Baek:2020owl}. Furthermore, we investigated four characteristic show-cases for the multi-component dark matter scenario. In order to fully understand the model and its phenomenology, an intensive parameter scan is better to be performed. We shall tackle them in future works.

\appendix

\section{Three-body Decay of $X_R$}
\label{apdx:XRdecay}
Consider a decay of particle 0 to particles 1, 2, and 3:
\begin{align*}
0 \rightarrow 1, 2, 3\,.
\end{align*}
The evolution of the number density of particle 0 is then governed by the following Boltzmann equation:
\begin{align}
\frac{dn_0}{dt} + 3Hn_0 &=
\int d\Pi_0 d\Pi_1 d\Pi_2 d\Pi_3 (2\pi)^4 \delta^{(4)}(p_0-p_1-p_2-p_3)
\nonumber\\
&\quad\times\left[
f_1 f_2 f_3 (1\pm f_0) |\mathcal{M}|^2_{1,2,3\rightarrow 0}  
-f_0(1\pm f_1)(1\pm f_2)(1\pm f_3)|\mathcal{M}|^2_{0\rightarrow 1,2,3}
\right]
\nonumber\\
&\approx
\int d\Pi_0 d\Pi_1 d\Pi_2 d\Pi_3 (2\pi)^4 \delta^{(4)}(p_0-p_1-p_2-p_3)|\mathcal{M}|^2_{0\rightarrow 1,2,3}
\nonumber\\
&\quad\times\left(
f_1 f_2 f_3 - f_0
\right)\,,
\end{align}
where
\begin{align}
d\Pi_i \equiv \frac{d^3p_i}{(2\pi)^3 2E_i}\,,
\end{align}
and we assumed $1\pm f_i \approx 1$ and CP conservation, i.e. $|\mathcal{M}|^2_{0\rightarrow 1,2,3}=|\mathcal{M}|^2_{1,2,3\rightarrow 0}$.
Now, using $f_i = (n_i/n_i^{\rm eq})f_i^{\rm eq}$ and $f_0^{\rm eq}=f_1^{\rm eq}f_2^{\rm eq}f_3^{\rm eq}$, we obtain
\begin{align}
\frac{dn_0}{dt}+3Hn_0 &\approx - \int d\Pi_0 d\Pi_1 d\Pi_2 d\Pi_3 (2\pi)^4 \delta^{(4)}(p_0-p_1-p_2-p_3)|\mathcal{M}|^2_{0\rightarrow 1,2,3}
\nonumber\\
&\quad\times \frac{f_0^{\rm eq}}{n_0^{\rm eq}}\left(
n_0 - n_0^{\rm eq}\frac{n_1 n_2 n_3}{n_1^{\rm eq}n_2^{\rm eq}n_3^{\rm eq}}
\right)\,.
\end{align}
Defining
\begin{align}
\Gamma_{0\rightarrow 1,2,3} \equiv
\frac{1}{2m_0}\int d\Pi_1d\Pi_2d\Pi_3(2\pi)^4\delta^{(4)}(p_0-p_1-p_2-p_3)
|\mathcal{M}|^2_{0\rightarrow 1, 2, 3}\,,
\end{align}
and the thermally-averaged decay rate
\begin{align}
\langle \Gamma \rangle_{0\rightarrow 1,2,3} \equiv
\int d\Pi_0 2m_0 \Gamma_{0\rightarrow 1,2,3}
\frac{f_0^{\rm eq}}{n_0^{\rm eq}}
\,,
\end{align}
the Boltzmann equation can be written as
\begin{align}
\frac{dn_0}{dt}+3Hn_0 = -\langle \Gamma \rangle_{0\rightarrow 1,2,3}\left(
n_0 - n_0^{\rm eq}\frac{n_1 n_2 n_3}{n_1^{\rm eq}n_2^{\rm eq}n_3^{\rm eq}}
\right)\,.
\end{align}
In terms of the yields, the Boltzmann equation can be rewritten as follows:
\begin{align}
\frac{dY_0}{dx} = -\frac{x}{H(m)}
\langle \Gamma \rangle_{0\rightarrow 1,2,3}\left(
Y_0 - Y_0^{\rm eq}\frac{Y_1Y_2Y_3}{Y_1^{\rm eq}Y_2^{\rm eq}Y_3^{\rm eq}}
\right)\,,
\end{align}
where $x \equiv m/T$ with $m$ being a mass variable.

For the decay process $X_R \rightarrow X_I+Y+Y^*$, we have
\begin{align}
\frac{dY_{X_R}}{dx}=-\frac{x}{H(m)}
\langle \Gamma \rangle_{X_R\rightarrow X_I,Y,Y^*}
\left(
Y_{X_R} - Y_{X_R}^{\rm eq}\frac{Y_{X_I}Y_y^2}{Y_{X_I}^{\rm eq}(Y_y^{\rm eq})^2}
\right)\,,
\end{align}
where $Y_y =  Y_Y + Y_{Y^*} = 2Y_Y$.
The decay rate is given by
\begin{align}
\langle\Gamma\rangle_{X_R\rightarrow X_I,Y,Y^*} &=
\frac{1}{64\pi^3 m_{X_R}} \int_{E_1^{\rm min}}^{E_1^{\rm max}} \int_{E_2^{\rm min}}^{E_2^{\rm max}}
|\mathcal{M}|^2_{X_R\rightarrow X_I,Y,Y^*} dE_2 dE_1
\,,\\
|\mathcal{M}|^2_{X_R\rightarrow X_I,Y,Y^*}&=
\frac{g_X^4 m_{X_R}^2 (E_1 + 2E_2 - m_{X_R})^2}{9(m_{X_I}^2 - 2m_{X_R}E_1 + m_{X_R}^2 - m_{Z'}^2)^2}
\,,
\end{align}
with
\begin{align}
E_{2}^{\rm min} &=
\frac{1}{2m_{23}^2}\left[
(m_{X_R}-E_1)m_{23}^2
-\sqrt{(E_1^2-m_{X_I}^2)\lambda(m_{23}^2,m_Y^2,m_Y^2)}
\right]
\,,\\
E_{2}^{\rm max} &=
\frac{1}{2m_{23}^2}\left[
(m_{X_R}-E_1)m_{23}^2
+\sqrt{(E_1^2-m_{X_I}^2)\lambda(m_{23}^2,m_Y^2,m_Y^2)}
\right]
\,,
\end{align}
and
\begin{align}
E_1^{\rm min} = m_{X_I} \,,\qquad
E_1^{\rm max} =
\frac{m_{X_R}^2 + m_{X_I}^2 - 4m_Y^2}{2m_{X_R}}\,,
\end{align}
where
\begin{align}
m_{23}^2 &= m_{X_R}^2 - 2m_{X_R}E_1 + m_{X_I}^2
\,,\\
\lambda(x,y,z) &\equiv
x^2+y^2+z^2-2xy-2xz-2yz
\,.
\end{align}

\section{The Full Boltzmann Equations}
\label{apdx:BoltzmannEqs}
In this appendix, we present the full Boltzmann equations for $Y$, $X_R$, $X_I$, and $Z^\prime$.
The Boltzmann equation for $Y$ is, using $Y_Y = Y_{Y^*} = Y_y/2$, given by
{\scriptsize
\begin{align}
\frac{dY_y}{dx} &=
\frac{2x}{H}\left[
\langle \Gamma \rangle_{Z^\prime\rightarrow Y,Y^*}\left(
Y_{Z^\prime} 
- \frac{Y_{Z^\prime}^{\rm eq}}{(Y_y^{\rm eq})^2}
Y_y^2
\right)
+
\langle \Gamma \rangle_{X_R\rightarrow X_I,Y,Y^*}\left(
Y_{X_R} 
- \frac{Y_{X_R}^{\rm eq}}{Y_{X_I}^{\rm eq}(Y_y^{\rm eq})^2}
Y_{X_I}Y_y^2
\right)
\right]
\nonumber\\
&\quad
+\frac{2s}{Hx^2}\Bigg[
\langle \sigma v \rangle_{Z',Z'\rightarrow Y,Y^*}\left(
Y_{Z'}^2
-\frac{(Y_{Z'}^{\rm eq})^2}{(Y_y^{\rm eq})^2}
Y_y^2
\right)
+\langle \sigma v \rangle_{X_R,X_R\rightarrow Y,Y^*}\left(
Y_{X_R}^2
-\frac{(Y_{X_R}^{\rm eq})^2}{(Y_y^{\rm eq})^2}
Y_y^2
\right)
\nonumber\\
&\quad
+\langle \sigma v \rangle_{X_I,X_I\rightarrow Y,Y^*}\left(
Y_{X_I}^2
-\frac{(Y_{X_I}^{\rm eq})^2}{(Y_y^{\rm eq})^2}
Y_y^2
\right)
+\langle \sigma v \rangle_{X_I,X_R\rightarrow Y,Y^*}\left(
Y_{X_I}Y_{X_R}
-\frac{Y_{X_I}^{\rm eq}Y_{X_R}^{\rm eq}}{(Y_y^{\rm eq})^2}
Y_y^2
\right)
\nonumber\\
&\quad
+\frac{1}{2}\langle \sigma v \rangle_{Z',Y\rightarrow Y^*,Y^*}\left(
Y_{Z'}Y_y
-\frac{Y_{Z'}^{\rm eq}}{Y_y^{\rm eq}}
Y_y^2
\right)
\Bigg]
\nonumber\\
&\quad
+\frac{2s^2}{Hx^5}\Bigg[
-\frac{1}{8}\langle \sigma v^2 \rangle_{Y,Y,Y^*\rightarrow Y^*,Y^*}\left(
Y_y^3
-Y_y^{\rm eq}
Y_y^2
\right)
-\frac{1}{8}\langle \sigma v^2 \rangle_{Y,Y,Y\rightarrow Y,Y^*}\left(
Y_y^3
-Y_y^{\rm eq}
Y_y^2
\right)
\nonumber\\
&\quad
-\frac{1}{4}\langle \sigma v^2 \rangle_{Z',Y,Y^* \rightarrow X_I, X_R}\left(
Y_{Z'}Y_y^2
-\frac{Y_{Z'}^{\rm eq}(Y_y^{\rm eq})^2}{Y_{X_I}^{\rm eq}Y_{X_R}^{\rm eq}}
Y_{X_I}Y_{X_R}
\right)
-\frac{1}{4}\langle \sigma v^2 \rangle_{X_I, Y, Y \rightarrow X_I, Y^*}\left(
Y_{X_I} Y_y^2
-Y_y^{\rm eq}
Y_{X_I} Y_y
\right)
\nonumber\\
&\quad
-\frac{3}{8}\langle \sigma v^2 \rangle_{Y,Y,Y\rightarrow X_I,X_R}\left(
Y_y^3
-\frac{(Y_y^{\rm eq})^3}{Y_{X_I}^{\rm eq}Y_{X_R}^{\rm eq}}
Y_{X_I}Y_{X_R}
\right)
-\frac{1}{4}\langle \sigma v^2 \rangle_{X_I,Y,Y\rightarrow X_R,Y^*}\left(
Y_{X_I}Y_y^2
-\frac{Y_{X_I}^{\rm eq}Y_y^{\rm eq}}{Y_{X_R}^{\rm eq}}
Y_{X_R}Y_y
\right)
\nonumber\\
&\quad
+\frac{1}{2}\langle \sigma v^2 \rangle_{X_I,X_I,Y\rightarrow Y^*,Y^*}\left(
Y_{X_I}^2Y_y
-\frac{(Y_{X_I}^{\rm eq})^2}{Y_y^{\rm eq}}
Y_y^2
\right)
-\frac{1}{4}\langle \sigma v^2 \rangle_{X_R,Y,Y^*\rightarrow Z',X_I}\left(
Y_{X_R}Y_y^2
-\frac{Y_{X_R}^{\rm eq}(Y_y^{\rm eq})^2}{Y_{Z'}^{\rm eq}Y_{X_I}^{\rm eq}}
Y_{Z'}Y_{X_I}
\right)
\nonumber\\
&\quad
-\frac{1}{4}\langle \sigma v^2 \rangle_{X_R,Y,Y\rightarrow X_I,Y^*}\left(
Y_{X_R}Y_y^2
-\frac{Y_{X_R}^{\rm eq}Y_y^{\rm eq}}{Y_{X_I}^{\rm eq}}
Y_{X_I}Y_y
\right)
-\frac{3}{8}\langle \sigma v^2 \rangle_{Y,Y,Y\rightarrow X_R,X_R}\left(
Y_y^3
-\frac{(Y_y^{\rm eq})^3}{(Y_{X_R}^{\rm eq})^2}
Y_{X_R}^2
\right)
\nonumber\\
&\quad
-\frac{1}{4}\langle \sigma v^2 \rangle_{X_R,Y,Y\rightarrow X_R,Y^*}\left(
Y_{X_R}Y_y^2
-Y_y^{\rm eq}
Y_{X_R}Y_y
\right)
+\frac{1}{2}\langle \sigma v^2 \rangle_{X_I,X_R,Y\rightarrow Y^*,Y^*}\left(
Y_{X_I}Y_{X_R}Y_y
-\frac{Y_{X_I}^{\rm eq}Y_{X_R}^{\rm eq}}{Y_y^{\rm eq}}
Y_y^2
\right)
\nonumber\\
&\quad
-\frac{1}{4}\langle \sigma v^2 \rangle_{Z',Y,Y^*\rightarrow X_I,X_I}\left(
Y_{Z'}Y_y^2
-\frac{Y_{Z'}^{\rm eq}(Y_y^{\rm eq})^2}{(Y_{X_I}^{\rm eq})^2}
Y_{X_I}^2
\right)
-\frac{3}{8}\langle \sigma v^2 \rangle_{Y,Y,Y\rightarrow X_I,X_I}\left(
Y_y^3
-\frac{(Y_y^{\rm eq})^3}{(Y_{X_I}^{\rm eq})^2}
Y_{X_I}^2
\right)
\nonumber\\
&\quad
+\frac{1}{2}\langle \sigma v^2 \rangle_{Z',Z',Y^*\rightarrow Y,Y}\left(
Y_{Z'}^2Y_y
-\frac{(Y_{Z'}^{\rm eq})^2}{Y_y^{\rm eq}}
Y_y^2
\right)
+\frac{1}{2}\langle \sigma v^2 \rangle_{X_R,X_R,Y\rightarrow Y^*,Y^*}\left(
Y_{X_R}^2 Y_y
-\frac{(Y_{X_R}^{\rm eq})^2}{Y_y^{\rm eq}}
Y_y^2
\right)
\nonumber\\
&\quad
-\frac{1}{4}\langle \sigma v^2 \rangle_{Z',Y^*,Y^*\rightarrow Z',Y}\left(
Y_{Z'}Y_y^2
-Y_y^{\rm eq}
Y_{Z'}Y_y
\right)
-\frac{1}{4}\langle \sigma v^2 \rangle_{Z',Y,Y^* \rightarrow X_R,X_R}\left(
Y_{Z'}Y_y^2
-\frac{Y_{Z'}^{\rm eq}(Y_y^{\rm eq})^2}{(Y_{X_R}^{\rm eq})^2}
Y_{X_R}^2
\right)
\nonumber\\
&\quad
+\langle \sigma v^2 \rangle_{Z',Z',Z'\rightarrow Y,Y^*}\left(
Y_{Z'}^3
-\frac{(Y_{Z'}^{\rm eq})^3}{(Y_y^{\rm eq})^2}
Y_y^2
\right)
+\langle \sigma v^2 \rangle_{Z',X_I,X_R\rightarrow Y,Y^*}\left(
Y_{Z'}Y_{X_I}Y_{X_R}
-\frac{Y_{Z'}^{\rm eq}Y_{X_I}^{\rm eq}Y_{X_R}^{\rm eq}}{(Y_y^{\rm eq})^2}
Y_y^2
\right)
\nonumber\\
&\quad
+\langle \sigma v^2 \rangle_{Z',X_R,X_R\rightarrow Y,Y^*}\left(
Y_{Z'}Y_{X_R}^2
-\frac{Y_{Z'}^{\rm eq}(Y_{X_R}^{\rm eq})^2}{(Y_y^{\rm eq})^2}
Y_y^2
\right)
+\langle \sigma v^2 \rangle_{Z',X_I,X_I\rightarrow Y,Y^*}\left(
Y_{Z'}Y_{X_I}^2
-\frac{Y_{Z'}^{\rm eq}(Y_{X_I}^{\rm eq})^2}{(Y_y^{\rm eq})^2}
Y_y^2
\right)
\Bigg]
\,.
\end{align}
}

The Boltzmann equations for $X_I$ and $X_R$ are
{\scriptsize
\begin{align}
\frac{dY_{X_I}}{dx} &=
\frac{x}{H}\Bigg[
\langle \Gamma \rangle_{X_R \rightarrow Z',X_I}\left(
Y_{X_R}
-\frac{Y_{X_R}^{\rm eq}}{Y_{Z'}^{\rm eq}Y_{X_I}^{\rm eq}}
Y_{Z'}Y_{X_I}
\right)
+\langle \Gamma \rangle_{Z' \rightarrow X_I,X_R}\left(
Y_{Z'}
-\frac{Y_{Z'}^{\rm eq}}{Y_{X_I}^{\rm eq}Y_{X_R}^{\rm eq}}
Y_{X_I}Y_{X_R}
\right)
\nonumber\\
&\quad
+
\langle \Gamma \rangle_{X_R\rightarrow X_I,Y,Y^*}\left(
Y_{X_R} 
- \frac{Y_{X_R}^{\rm eq}}{Y_{X_I}^{\rm eq}(Y_y^{\rm eq})^2}
Y_{X_I}Y_y^2
\right)
\Bigg]
\nonumber\\ 
&\quad
+\frac{s}{Hx^2}\Bigg[
2\langle \sigma v \rangle_{Z',Z'\rightarrow X_I,X_I}\left(
Y_{Z'}^2
-\frac{(Y_{Z'}^{\rm eq})^2}{(Y_{X_I}^{\rm eq})^2}
Y_{X_I}^2
\right)
+2\langle \sigma v \rangle_{X_R, X_R\rightarrow X_I,X_I}\left(
Y_{X_R}^2
-\frac{(Y_{X_R}^{\rm eq})^2}{(Y_{X_I}^{\rm eq})^2}
Y_{X_I}^2
\right)
\nonumber\\
&\quad
+\langle \sigma v \rangle_{X_R,Y\rightarrow X_I,Y}\left(
Y_{X_R}Y_y
-\frac{Y_{X_R}^{\rm eq}}{Y_{X_I}^{\rm eq}}
Y_{X_I}Y_y
\right)
-2\langle \sigma v \rangle_{X_I,X_I \rightarrow Z',Z'}\left(
Y_{X_I}^2
-\frac{(Y_{X_I}^{\rm eq})^2}{(Y_{Z'}^{\rm eq})^2}
Y_{Z'}^2
\right)
\nonumber\\
&\quad
-2\langle \sigma v \rangle_{X_I,X_I \rightarrow Y,Y^*}\left(
Y_{X_I}^2
-\frac{(Y_{X_I}^{\rm eq})^2}{(Y_{y}^{\rm eq})^2}
Y_{y}^2
\right)
-\langle \sigma v \rangle_{X_I,X_R \rightarrow Y,Y^*}\left(
Y_{X_I}Y_{X_R}
-\frac{Y_{X_I}^{\rm eq}Y_{X_R}^{\rm eq}}{(Y_{y}^{\rm eq})^2}
Y_{y}^2
\right)
\Bigg]
\nonumber\\ 
&\quad
+\frac{s^2}{Hx^5}\Bigg[
-\langle \sigma v^2 \rangle_{Z',X_I,X_R\rightarrow Z',Z'}\left(
Y_{Z'}Y_{X_I}Y_{X_R}
-\frac{Y_{X_I}^{\rm eq}Y_{X_R}^{\rm eq}}{Y_{Z'}^{\rm eq}}
Y_{Z'}^2
\right)
-\langle \sigma v^2 \rangle_{X_I,X_R,Y\rightarrow Z',Y}\left(
Y_{X_I}Y_{X_R}Y_y
-\frac{Y_{X_I}^{\rm eq}Y_{X_R}^{\rm eq}}{Y_{Z'}^{\rm eq}}
Y_{Z'}Y_y
\right)
\nonumber\\
&\quad
+\frac{1}{4}\langle \sigma v^2 \rangle_{Z',Y,Y^*\rightarrow X_I,X_R}\left(
Y_{Z'}Y_y^2
-\frac{Y_{Z'}^{\rm eq}(Y_{y}^{\rm eq})^2}{Y_{X_I}^{\rm eq}Y_{X_R}^{\rm eq}}
Y_{X_I}Y_{X_R}
\right)
+\frac{1}{4}\langle \sigma v^2 \rangle_{Y,Y,Y\rightarrow X_I,X_R}\left(
Y_y^3
-\frac{(Y_y^{\rm eq})^3}{Y_{X_I}^{\rm eq}Y_{X_R}^{\rm eq}}
Y_{X_I}Y_{X_R}
\right)
\nonumber\\
&\quad
-\frac{1}{2}\langle \sigma v^2 \rangle_{X_I,Y,Y\rightarrow X_R,Y^*}\left(
Y_{X_I}Y_y^2
-\frac{Y_{X_I}^{\rm eq}Y_y^{\rm eq}}{Y_{X_R}^{\rm eq}}
Y_{X_R}Y_y
\right)
-2\langle \sigma v^2 \rangle_{X_I,X_I,Y\rightarrow Y^*,Y^*}\left(
Y_{X_I}^2Y_y
-\frac{(Y_{X_I}^{\rm eq})^2}{Y_y^{\rm eq}}
Y_y^2
\right)
\nonumber\\
&\quad
+\langle \sigma v^2 \rangle_{Z',Z',X_R\rightarrow Z',X_I}\left(
Y_{Z'}^2Y_{X_R}
-\frac{Y_{Z'}^{\rm eq}Y_{X_R}^{\rm eq}}{Y_{X_I}^{\rm eq}}
Y_{Z'}Y_{X_I}
\right)
-\langle \sigma v^2 \rangle_{X_I,X_I,X_R\rightarrow Z',X_I}\left(
Y_{X_I}^2Y_{X_R}
-\frac{Y_{X_I}^{\rm eq}Y_{X_R}^{\rm eq}}{Y_{Z'}^{\rm eq}}
Y_{Z'}Y_{X_I}
\right)
\nonumber\\
&\quad
+\langle \sigma v^2 \rangle_{X_R,X_R,X_R\rightarrow Z',X_I}\left(
Y_{X_R}^3
-\frac{(Y_{X_R}^{\rm eq})^3}{Y_{Z'}^{\rm eq}Y_{X_I}^{\rm eq}}
Y_{Z'}Y_{X_I}
\right)
+\frac{1}{4}\langle \sigma v^2 \rangle_{X_R,Y,Y^*\rightarrow Z',X_I}\left(
Y_{X_R}Y_y^2
-\frac{Y_{X_R}^{\rm eq}(Y_y^{\rm eq})^2}{Y_{Z'}^{\rm eq}Y_{X_I}^{\rm eq}}
Y_{Z'}Y_{X_I}
\right)
\nonumber\\
&\quad
+\langle \sigma v^2 \rangle_{Z',X_I,X_R\rightarrow X_I,X_I}\left(
Y_{Z'}Y_{X_I}Y_{X_R}
-\frac{Y_{Z'}^{\rm eq}Y_{X_R}^{\rm eq}}{Y_{X_I}^{\rm eq}}
Y_{X_I}^2
\right)
+\frac{1}{2}\langle \sigma v^2 \rangle_{X_R,Y,Y\rightarrow X_I,Y^*}\left(
Y_{X_R}Y_y^2
-\frac{Y_{X_R}^{\rm eq}Y_y^{\rm eq}}{Y_{X_I}^{\rm eq}}
Y_{X_I}Y_y
\right)
\nonumber\\
&\quad
-\langle \sigma v^2 \rangle_{X_I,X_R,Y\rightarrow Y^*,Y^*}\left(
Y_{X_I}Y_{X_R}Y_y
-\frac{Y_{X_I}^{\rm eq}Y_{X_R}^{\rm eq}}{Y_y^{\rm eq}}
Y_y^2
\right)
-\langle \sigma v^2 \rangle_{Z',Z',X_I\rightarrow Z',X_R}\left(
Y_{Z'}^2Y_{X_I}
-\frac{Y_{Z'}^{\rm eq}Y_{X_I}^{\rm eq}}{Y_{X_R}^{\rm eq}}
Y_{Z'}Y_{X_R}
\right)
\nonumber\\
&\quad
-3\langle \sigma v^2 \rangle_{X_I,X_I,X_I\rightarrow Z',X_R}\left(
Y_{X_I}^3
-\frac{(Y_{X_I}^{\rm eq})^3}{Y_{Z'}^{\rm eq}Y_{X_R}^{\rm eq}}
Y_{Z'}Y_{X_R}
\right)
-\langle \sigma v^2 \rangle_{X_I,X_R,X_R\rightarrow Z',X_R}\left(
Y_{X_I}Y_{X_R}^2
-\frac{Y_{X_I}^{\rm eq}Y_{X_R}^{\rm eq}}{Y_{Z'}^{\rm eq}}
Y_{Z'}Y_{X_R}
\right)
\nonumber\\
&\quad
+\frac{1}{2}\langle \sigma v^2 \rangle_{Z',Y,Y^* \rightarrow X_I,X_I}\left(
Y_{Z'}Y_y^2
-\frac{Y_{Z'}^{\rm eq}(Y_y^{\rm eq})^2}{(Y_{X_I}^{\rm eq})^2}
Y_{X_I}^2
\right)
+\frac{1}{2}\langle \sigma v^2 \rangle_{Y,Y,Y\rightarrow X_I,X_I}\left(
Y_y^3
-\frac{(Y_y^{\rm eq})^3}{(Y_{X_I}^{\rm eq})^2}
Y_{X_I}^2
\right)
\nonumber\\
&\quad
-\langle \sigma v^2 \rangle_{Z',X_I,X_R\rightarrow X_R,X_R}\left(
Y_{Z'}Y_{X_I}Y_{X_R}
-\frac{Y_{Z'}^{\rm eq}Y_{X_I}^{\rm eq}}{Y_{X_R}^{\rm eq}}
Y_{X_R}^2
\right)
-\langle \sigma v^2 \rangle_{Z',X_I,Y\rightarrow X_R,Y}\left(
Y_{Z'}Y_{X_I}Y_y
-\frac{Y_{Z'}^{\rm eq}Y_{X_I}^{\rm eq}}{Y_{X_R}^{\rm eq}}
Y_{X_R}Y_y
\right)
\nonumber\\
&\quad
-2\langle \sigma v^2 \rangle_{X_I,X_I,Y\rightarrow Z',Y}\left(
Y_{X_I}^2Y_y
-\frac{(Y_{X_I}^{\rm eq})^2}{Y_{Z'}^{\rm eq}}
Y_{Z'}Y_y
\right)
+\langle \sigma v^2 \rangle_{Z',Z',Z'\rightarrow X_I,X_R}\left(
Y_{Z'}^3
-\frac{(Y_{Z'}^{\rm eq})^3}{Y_{X_I}^{\rm eq}Y_{X_R}^{\rm eq}}
Y_{X_I}Y_{X_R}
\right)
\nonumber\\
&\quad
-\langle \sigma v^2 \rangle_{Z',X_I,X_I\rightarrow X_I,X_R}\left(
Y_{Z'}Y_{X_I}^2
-\frac{Y_{Z'}^{\rm eq}Y_{X_I}^{\rm eq}}{Y_{X_R}^{\rm eq}}
Y_{X_I}Y_{X_R}
\right)
+\langle \sigma v^2 \rangle_{Z',X_R,X_R\rightarrow X_I,X_R}\left(
Y_{Z'}Y_{X_R}^2
-\frac{Y_{Z'}^{\rm eq}Y_{X_R}^{\rm eq}}{Y_{X_I}^{\rm eq}}
Y_{X_I}Y_{X_R}
\right)
\nonumber\\
&\quad
+\langle \sigma v^2 \rangle_{Z',X_R,Y\rightarrow X_I,Y}\left(
Y_{Z'}Y_{X_R}Y_y
-\frac{Y_{Z'}^{\rm eq}Y_{X_R}^{\rm eq}}{Y_{X_I}^{\rm eq}}
Y_{X_I}Y_y
\right)
-\langle \sigma v^2 \rangle_{Z',X_I,X_R\rightarrow Y,Y^*}\left(
Y_{Z'}Y_{X_I}Y_{X_R}
-\frac{Y_{Z'}^{\rm eq}Y_{X_I}^{\rm eq}Y_{X_R}^{\rm eq}}{(Y_y^{\rm eq})^2}
Y_y^2
\right)
\nonumber\\
&\quad
-2\langle \sigma v^2 \rangle_{Z',X_I,X_I\rightarrow Y,Y^*}\left(
Y_{Z'}(Y_{X_I})^2
-\frac{Y_{Z'}^{\rm eq}(Y_{X_I}^{\rm eq})^2}{(Y_y^{\rm eq})^2}
Y_y^2
\right)
\Bigg]
\,,
\end{align}
}
and
{\scriptsize
\begin{align}
\frac{dY_{X_R}}{dx} &=
\frac{x}{H}\Bigg[
\langle \Gamma \rangle_{Z'\rightarrow X_I,X_R}\left(
Y_{Z'}
-\frac{Y_{Z'}^{\rm eq}}{Y_{X_I}^{\rm eq}Y_{X_R}^{\rm eq}}
Y_{X_I}Y_{X_R}
\right)
-\langle \Gamma \rangle_{X_R \rightarrow Z',X_I}\left(
Y_{X_R}
-\frac{Y_{X_R}^{\rm eq}}{Y_{Z'}^{\rm eq}Y_{X_I}^{\rm eq}}
Y_{Z'}Y_{X_I}
\right)
\nonumber\\
&\quad
-
\langle \Gamma \rangle_{X_R\rightarrow X_I,Y,Y^*}\left(
Y_{X_R} 
- \frac{Y_{X_R}^{\rm eq}}{Y_{X_I}^{\rm eq}(Y_y^{\rm eq})^2}
Y_{X_I}Y_y^2
\right)
\Bigg]
\nonumber\\ 
&\quad
+\frac{s}{Hx^2}\Bigg[
2\langle \sigma v \rangle_{Z',Z'\rightarrow X_R,X_R}\left(
Y_{Z'}^2
-\frac{(Y_{Z'}^{\rm eq})^2}{(Y_{X_R}^{\rm eq})^2}
Y_{X_R}^2
\right)
-2\langle \sigma v \rangle_{X_R,X_R \rightarrow Z',Z'}\left(
Y_{X_R}^2
-\frac{(Y_{X_R}^{\rm eq})^2}{(Y_{Z'}^{\rm eq})^2}
Y_{Z'}^2
\right)
\nonumber\\
&\quad
-2\langle \sigma v \rangle_{X_R,X_R\rightarrow X_I,X_I}\left(
Y_{X_R}^2
-\frac{(Y_{X_R}^{\rm eq})^2}{(Y_{X_I}^{\rm eq})^2}
Y_{X_I}^2
\right)
-2\langle \sigma v \rangle_{X_R,X_R\rightarrow Y,Y^*}\left(
Y_{X_R}^2
-\frac{(Y_{X_R}^{\rm eq})^2}{(Y_y^{\rm eq})^2}
Y_y^2
\right)
\nonumber\\
&\quad
-\langle \sigma v \rangle_{X_R,Y\rightarrow X_I,Y}\left(
Y_{X_R}Y_y
-\frac{Y_{X_R}^{\rm eq}}{Y_{X_I}^{\rm eq}}
Y_{X_I}Y_y
\right)
-\langle \sigma v \rangle_{X_I,X_R \rightarrow Y,Y^*}\left(
Y_{X_I}Y_{X_R}
-\frac{Y_{X_I}^{\rm eq}Y_{X_R}^{\rm eq}}{(Y_y^{\rm eq})^2}
Y_y^2
\right)
\Bigg]
\nonumber\\ 
&\quad
+\frac{s^2}{Hx^5}\Bigg[
-\langle \sigma v^2 \rangle_{Z',X_I,X_R\rightarrow Z',Z'}\left(
Y_{Z'}Y_{X_I}Y_{X_R}
-\frac{Y_{X_I}^{\rm eq}Y_{X_R}^{\rm eq}}{Y_{Z'}^{\rm eq}}
Y_{Z'}^2
\right)
-\langle \sigma v^2 \rangle_{X_I,X_R,Y\rightarrow Z',Y}\left(
Y_{X_I}Y_{X_R}Y_y
-\frac{Y_{X_I}^{\rm eq}Y_{X_R}^{\rm eq}}{Y_{Z'}^{\rm eq}}
Y_{Z'}Y_y
\right)
\nonumber\\
&\quad
-2\langle \sigma v^2 \rangle_{X_R,X_R,Y\rightarrow Z',Y}\left(
Y_{X_R}^2Y_y
-\frac{(Y_{X_R}^{\rm eq})^2}{Y_{Z'}^{\rm eq}}
Y_{Z'}Y_y
\right)
+\frac{1}{4}\langle \sigma v^2 \rangle_{Z',Y,Y^*\rightarrow X_I,X_R}\left(
Y_{Z'}Y_y^2
-\frac{Y_{Z'}^{\rm eq}(Y_y^{\rm eq})^2}{Y_{X_I}^{\rm eq}Y_{X_R}^{\rm eq}}
Y_{X_I}Y_{X_R}
\right)
\nonumber\\
&\quad
+\frac{1}{4}\langle \sigma v^2 \rangle_{Y,Y,Y\rightarrow X_I,X_R}\left(
Y_y^3
-\frac{(Y_y^{\rm eq})^3}{Y_{X_I}^{\rm eq}Y_{X_R}^{\rm eq}}
Y_{X_I}Y_{X_R}
\right)
+\frac{1}{2}\langle \sigma v^2 \rangle_{X_I,Y^*,Y^*\rightarrow X_R,Y}\left(
Y_{X_I}Y_y^2
-\frac{Y_{X_I}^{\rm eq}Y_y^{\rm eq}}{Y_{X_R}^{\rm eq}}
Y_{X_R}Y_y
\right)
\nonumber\\
&\quad
-\langle \sigma v^2 \rangle_{Z',Z',X_R\rightarrow Z',X_I}\left(
Y_{Z'}^2 Y_{X_R}
-\frac{Y_{Z'}^{\rm eq}Y_{X_R}^{\rm eq}}{Y_{X_I}^{\rm eq}}
Y_{Z'}Y_{X_I}
\right)
-\langle \sigma v^2 \rangle_{X_I,X_I,X_R\rightarrow Z',X_I}\left(
Y_{X_I}^2 Y_{X_R}
-\frac{Y_{X_I}^{\rm eq}Y_{X_R}^{\rm eq}}{Y_{Z'}^{\rm eq}}
Y_{Z'}Y_{X_I}
\right)
\nonumber\\
&\quad
-3\langle \sigma v^2 \rangle_{X_R,X_R,X_R\rightarrow Z',X_I}\left(
Y_{X_R}^3
-\frac{(Y_{X_R}^{\rm eq})^3}{Y_{Z'}^{\rm eq}Y_{X_I}^{\rm eq}}
Y_{Z'}Y_{X_I}
\right)
-\frac{1}{4}\langle \sigma v^2 \rangle_{X_R,Y,Y^*\rightarrow Z',X_I}\left(
Y_{X_R}Y_y^2
-\frac{Y_{X_R}^{\rm eq}(Y_y^{\rm eq})^2}{Y_{Z'}^{\rm eq}Y_{X_I}^{\rm eq}}
Y_{Z'}Y_{X_I}
\right)
\nonumber\\
&\quad
-\langle \sigma v^2 \rangle_{Z',X_I,X_R\rightarrow X_I,X_I}\left(
Y_{Z'}Y_{X_I}Y_{X_R}
-\frac{Y_{Z'}^{\rm eq}Y_{X_R}^{\rm eq}}{Y_{X_I}^{\rm eq}}
Y_{X_I}^2
\right)
-\frac{1}{2}\langle \sigma v^2 \rangle_{X_R,Y,Y\rightarrow X_I,Y^*}\left(
Y_{X_R} Y_y^2
-\frac{Y_{X_R}^{\rm eq}Y_y^{\rm eq}}{Y_{X_I}^{\rm eq}}
Y_{X_I}Y_y
\right)
\nonumber\\
&\quad
+\frac{1}{2}\langle \sigma v^2 \rangle_{Y,Y,Y\rightarrow X_R,X_R}\left(
Y_y^3
-\frac{(Y_y^{\rm eq})^3}{(Y_{X_R}^{\rm eq})^2}
Y_{X_R}^2
\right)
-\langle \sigma v^2 \rangle_{X_I,X_R,Y^*\rightarrow Y,Y}\left(
Y_{X_I}Y_{X_R}Y_y
-\frac{Y_{X_I}^{\rm eq}Y_{X_R}^{\rm eq}}{Y_y^{\rm eq}}
Y_y^2
\right)
\nonumber\\
&\quad
+\langle \sigma v^2 \rangle_{Z',Z',X_I\rightarrow Z',X_R}\left(
Y_{Z'}^2Y_{X_I}
-\frac{Y_{Z'}^{\rm eq}Y_{X_I}^{\rm eq}}{Y_{X_R}^{\rm eq}}
Y_{Z'}Y_{X_R}
\right)
+\langle \sigma v^2 \rangle_{X_I,X_I,X_I\rightarrow Z',X_R}\left(
Y_{X_I}^3
-\frac{(Y_{X_I}^{\rm eq})^3}{Y_{Z'}^{\rm eq}Y_{X_R}^{\rm eq}}
Y_{Z'}Y_{X_R}
\right)
\nonumber\\
&\quad
-\langle \sigma v^2 \rangle_{X_I,X_R,X_R\rightarrow Z',X_R}\left(
Y_{X_I}Y_{X_R}^2
-\frac{Y_{X_I}^{\rm eq}Y_{X_R}^{\rm eq}}{Y_{Z'}^{\rm eq}}
Y_{Z'}Y_{X_R}
\right)
+\langle \sigma v^2 \rangle_{Z',X_I,X_R\rightarrow X_R,X_R}\left(
Y_{Z'}Y_{X_I}Y_{X_R}
-\frac{Y_{Z'}^{\rm eq}Y_{X_I}^{\rm eq}}{Y_{X_R}^{\rm eq}}
Y_{X_R}^2
\right)
\nonumber\\
&\quad
+\langle \sigma v^2 \rangle_{Z',X_I,Y\rightarrow X_R,Y}\left(
Y_{Z'}Y_{X_I}Y_y
-\frac{Y_{Z'}^{\rm eq}Y_{X_I}^{\rm eq}}{Y_{X_R}^{\rm eq}}
Y_{X_R}Y_y
\right)
-2\langle \sigma v^2 \rangle_{X_R,X_R,Y^*\rightarrow Y,Y}\left(
Y_{X_R}^2Y_y
-\frac{(Y_{X_R}^{\rm eq})^2}{Y_y^{\rm eq}}
Y_y^2
\right)
\nonumber\\
&\quad
+\langle \sigma v^2 \rangle_{Z',Z',Z'\rightarrow X_I,X_R}\left(
Y_{Z'}^3
-\frac{(Y_{Z'}^{\rm eq})^3}{Y_{X_I}^{\rm eq}Y_{X_R}^{\rm eq}}
Y_{X_I}Y_{X_R}
\right)
+\langle \sigma v^2 \rangle_{Z',X_I,X_I\rightarrow X_I,X_R}\left(
Y_{Z'}Y_{X_I}^2
-\frac{Y_{Z'}^{\rm eq}Y_{X_I}^{\rm eq}}{Y_{X_R}^{\rm eq}}
Y_{X_I}Y_{X_R}
\right)
\nonumber\\
&\quad
-\langle \sigma v^2 \rangle_{Z',X_R,X_R\rightarrow X_I,X_R}\left(
Y_{Z'}Y_{X_R}^2
-\frac{Y_{Z'}^{\rm eq}Y_{X_R}^{\rm eq}}{Y_{X_I}^{\rm eq}}
Y_{X_I}Y_{X_R}
\right)
-\langle \sigma v^2 \rangle_{Z',X_R,Y\rightarrow X_I,Y}\left(
Y_{Z'}Y_{X_R}Y_y
-\frac{Y_{Z'}^{\rm eq}Y_{X_R}^{\rm eq}}{Y_{X_I}^{\rm eq}}
Y_{X_I}Y_y
\right)
\nonumber\\
&\quad
+\frac{1}{2}\langle \sigma v^2 \rangle_{Z',Y,Y^*\rightarrow X_R,X_R}\left(
Y_{Z'}Y_y^2
-\frac{Y_{Z'}^{\rm eq}(Y_y^{\rm eq})^2}{(Y_{X_R}^{\rm eq})^2}
Y_{X_R}^2
\right)
-\langle \sigma v^2 \rangle_{Z',X_I,X_R\rightarrow Y,Y^*}\left(
Y_{Z'}Y_{X_I}Y_{X_R}
-\frac{Y_{Z'}^{\rm eq}Y_{X_I}^{\rm eq}Y_{X_R}^{\rm eq}}{(Y_y^{\rm eq})^2}
Y_y^2
\right)
\nonumber\\
&\quad
-2\langle \sigma v^2 \rangle_{Z',X_R,X_R\rightarrow Y,Y^*}\left(
Y_{Z'}Y_{X_R}^2
-\frac{Y_{Z'}^{\rm eq}(Y_{X_R}^{\rm eq})^2}{(Y_y^{\rm eq})^2}
Y_y^2
\right)
\Bigg]
\,.
\end{align}
}

Finally, the Boltzmann equation for $Z'$ is given by
{\scriptsize
\begin{align}
\frac{dY_{Z'}}{dx} &=
\frac{x}{H}\Bigg[
-\langle \Gamma \rangle_{Z'\rightarrow Y,Y^*}\left(
Y_{Z'}
-\frac{Y_{Z'}^{\rm eq}}{(Y_y^{\rm eq})^2}
Y_y^2
\right)
-\langle \Gamma \rangle_{Z'\rightarrow X_I,X_R}\left(
Y_{Z'}
-\frac{Y_{Z'}^{\rm eq}}{Y_{X_I}^{\rm eq}Y_{X_R}^{\rm eq}}
Y_{X_I}Y_{X_R}
\right)
\nonumber\\
&\quad
+\langle \Gamma \rangle_{X_R \rightarrow Z',X_I}\left(
Y_{X_R}
-\frac{Y_{X_R}^{\rm eq}}{Y_{Z'}^{\rm eq}Y_{X_I}^{\rm eq}}
Y_{Z'}Y_{X_I}
\right)
\Bigg]
\nonumber\\ 
&\quad
+\frac{s}{Hx^2}\Bigg[
-2\langle \sigma v \rangle_{Z',Z'\rightarrow X_I,X_I}\left(
Y_{Z'}^2
-\frac{(Y_{Z'}^{\rm eq})^2}{(Y_{X_I}^{\rm eq})^2}
Y_{X_I}^2
\right)
-2\langle \sigma v \rangle_{Z',Z'\rightarrow X_R,X_R}\left(
Y_{Z'}^2
-\frac{(Y_{Z'}^{\rm eq})^2}{(Y_{X_R}^{\rm eq})^2}
Y_{X_R}^2
\right)
\nonumber\\
&\quad
-2\langle \sigma v \rangle_{Z',Z'\rightarrow Y,Y^*}\left(
Y_{Z'}^2
-\frac{(Y_{Z'}^{\rm eq})^2}{(Y_y^{\rm eq})^2}
Y_y^2
\right)
+2\langle \sigma v \rangle_{X_R,X_R\rightarrow Z',Z'}\left(
Y_{X_R}^2
-\frac{(Y_{X_R}^{\rm eq})^2}{(Y_{Z'}^{\rm eq})^2}
Y_{Z'}^2
\right)
\nonumber\\
&\quad
-\langle \sigma v \rangle_{Z',Y\rightarrow Y^*,Y^*}\left(
Y_{Z'}Y_y
-\frac{Y_{Z'}^{\rm eq}}{Y_y^{\rm eq}}
Y_y^2
\right)
+2\langle \sigma v \rangle_{X_I,X_I\rightarrow Z',Z'}\left(
Y_{X_I}^2
-\frac{(Y_{X_I}^{\rm eq})^2}{(Y_{Z'}^{\rm eq})^2}
Y_{Z'}^2
\right)
\Bigg]
\nonumber\\ 
&\quad
+\frac{s^2}{Hx^5}\Bigg[
\langle \sigma v^2 \rangle_{Z',X_I,X_R\rightarrow Z',Z'}\left(
Y_{Z'}Y_{X_I}Y_{X_R}
-\frac{Y_{X_I}^{\rm eq}Y_{X_R}^{\rm eq}}{Y_{Z'}^{\rm eq}}
Y_{Z'}^2
\right)
+\langle \sigma v^2 \rangle_{X_I,X_R,Y\rightarrow Z',Y}\left(
Y_{X_I}Y_{X_R}Y_y
-\frac{Y_{X_I}^{\rm eq}Y_{X_R}^{\rm eq}}{Y_{Z'}^{\rm eq}}
Y_{Z'}Y_y
\right)
\nonumber\\
&\quad
+\langle \sigma v^2 \rangle_{X_R,X_R,Y\rightarrow Z',Y}\left(
Y_{X_R}^2Y_y
-\frac{(Y_{X_R}^{\rm eq})^2}{Y_{Z'}^{\rm eq}}
Y_{Z'}Y_y
\right)
-\frac{1}{4}\langle \sigma v^2 \rangle_{Z',Y,Y^*\rightarrow X_I,X_R}\left(
Y_{Z'}Y_y^2
-\frac{Y_{Z'}^{\rm eq}(Y_y^{\rm eq})^2}{Y_{X_I}^{\rm eq}Y_{X_R}^{\rm eq}}
Y_{X_I}Y_{X_R}
\right)
\nonumber\\
&\quad
-\langle \sigma v^2 \rangle_{Z',Z',X_R\rightarrow Z',X_I}\left(
Y_{Z'}^2Y_{X_R}
-\frac{Y_{Z'}^{\rm eq}Y_{X_R}^{\rm eq}}{Y_{X_I}^{\rm eq}}
Y_{Z'}Y_{X_I}
\right)
+\langle \sigma v^2 \rangle_{X_I,X_I,X_R\rightarrow Z',X_I}\left(
Y_{X_I}^2Y_{X_R}
-\frac{Y_{X_I}^{\rm eq}Y_{X_R}^{\rm eq}}{Y_{Z'}^{\rm eq}}
Y_{Z'}Y_{X_I}
\right)
\nonumber\\
&\quad
+\langle \sigma v^2 \rangle_{X_R,X_R,X_R\rightarrow Z',X_I}\left(
Y_{X_R}^3
-\frac{(Y_{X_R}^{\rm eq})^3}{Y_{Z'}^{\rm eq}Y_{X_I}^{\rm eq}}
Y_{Z'}Y_{X_I}
\right)
+\frac{1}{4}\langle \sigma v^2 \rangle_{X_R,Y,Y^*\rightarrow Z',X_I}\left(
Y_{X_R} Y_y^2
-\frac{Y_{X_R}^{\rm eq}(Y_y^{\rm eq})^2}{Y_{Z'}^{\rm eq}Y_{X_I}^{\rm eq}}
Y_{Z'}Y_{X_I}
\right)
\nonumber\\
&\quad
-\langle \sigma v^2 \rangle_{Z',X_I,X_R\rightarrow X_I,X_I}\left(
Y_{Z'}Y_{X_I}Y_{X_R}
-\frac{Y_{Z'}^{\rm eq}Y_{X_R}^{\rm eq}}{Y_{X_I}^{\rm eq}}
Y_{X_I}^2
\right)
-\langle \sigma v^2 \rangle_{Z',Z',X_I \rightarrow Z',X_R}\left(
Y_{Z'}^2Y_{X_I}
-\frac{Y_{Z'}^{\rm eq}Y_{X_I}^{\rm eq}}{Y_{X_R}^{\rm eq}}
Y_{Z'}Y_{X_R}
\right)
\nonumber\\
&\quad
+\langle \sigma v^2 \rangle_{X_I,X_I,X_I\rightarrow Z',X_R}\left(
Y_{X_I}^3
-\frac{(Y_{X_I}^{\rm eq})^3}{Y_{Z'}^{\rm eq}Y_{X_R}^{\rm eq}}
Y_{Z'}Y_{X_R}
\right)
+\langle \sigma v^2 \rangle_{X_I,X_R,X_R\rightarrow Z',X_R}\left(
Y_{X_I}Y_{X_R}^2
-\frac{Y_{X_I}^{\rm eq}Y_{X_R}^{\rm eq}}{Y_{Z'}^{\rm eq}}
Y_{Z'}Y_{X_R}
\right)
\nonumber\\
&\quad
-\langle \sigma v^2 \rangle_{Z',Z',Y\rightarrow Z',Y}\left(
Y_{Z'}^2Y_y
-Y_{Z'}^{\rm eq}
Y_{Z'}Y_y
\right)
-\frac{1}{4}\langle \sigma v^2 \rangle_{Z',Y,Y^*\rightarrow X_I,X_I}\left(
Y_{Z'}Y_y^2
-\frac{Y_{Z'}^{\rm eq}(Y_y^{\rm eq})^2}{(Y_{X_I}^{\rm eq})^2}
Y_{X_I}^2
\right)
\nonumber\\
&\quad
-\langle \sigma v^2 \rangle_{Z',X_I,Y\rightarrow X_I,Y}\left(
Y_{Z'}Y_{X_I}Y_y
-Y_{Z'}^{\rm eq}
Y_{X_I}Y_y
\right)
-\langle \sigma v^2 \rangle_{Z',X_I,X_R \rightarrow X_R,X_R}\left(
Y_{Z'}Y_{X_I}Y_{X_R}
-\frac{Y_{Z'}^{\rm eq}Y_{X_I}^{\rm eq}}{Y_{X_R}^{\rm eq}}
Y_{X_R}^2
\right)
\nonumber\\
&\quad
-\langle \sigma v^2 \rangle_{Z',X_I,Y\rightarrow X_R,Y}\left(
Y_{Z'}Y_{X_I}Y_y
-\frac{Y_{Z'}^{\rm eq}Y_{X_I}^{\rm eq}}{Y_{X_R}^{\rm eq}}
Y_{X_R}Y_y
\right)
-2\langle \sigma v^2 \rangle_{Z',Z',Y\rightarrow Y^*,Y^*}\left(
Y_{Z'}^2 Y_y
-\frac{(Y_{Z'}^{\rm eq})^2}{Y_y^{\rm eq}}
Y_y^2
\right)
\nonumber\\
&\quad
+\langle \sigma v^2 \rangle_{X_I,X_I,Y\rightarrow Z',Y}\left(
Y_{X_I}^2 Y_y
-\frac{(Y_{X_I}^{\rm eq})^2}{Y_{Z'}^{\rm eq}}
Y_{Z'}Y_y
\right)
-3\langle \sigma v^2 \rangle_{Z',Z',Z'\rightarrow X_I,X_R}\left(
Y_{Z'}^3
-\frac{(Y_{Z'}^{\rm eq})^3}{Y_{X_I}^{\rm eq}Y_{X_R}^{\rm eq}}
Y_{X_I}Y_{X_R}
\right)
\nonumber\\
&\quad
-\langle \sigma v^2 \rangle_{Z',X_I,X_I\rightarrow X_I,X_R}\left(
Y_{Z'}Y_{X_I}^2
-\frac{Y_{Z'}^{\rm eq}Y_{X_I}^{\rm eq}}{Y_{X_R}^{\rm eq}}
Y_{X_I}Y_{X_R}
\right)
-\langle \sigma v^2 \rangle_{Z',X_R,X_R\rightarrow X_I,X_R}\left(
Y_{Z'}Y_{X_R}^2
-\frac{Y_{Z'}^{\rm eq}Y_{X_R}^{\rm eq}}{Y_{X_I}^{\rm eq}}
Y_{X_I}Y_{X_R}
\right)
\nonumber\\
&\quad
-\langle \sigma v^2 \rangle_{Z',X_R,Y\rightarrow X_I,Y}\left(
Y_{Z'}Y_{X_R} Y_y
-\frac{Y_{Z'}^{\rm eq}Y_{X_R}^{\rm eq}}{Y_{X_I}^{\rm eq}}
Y_{X_I} Y_y
\right)
-\frac{1}{4}\langle \sigma v^2 \rangle_{Z',Y,Y^*\rightarrow X_R,X_R}\left(
Y_{Z'}Y_y^2
-\frac{Y_{Z'}^{\rm eq}(Y_y^{\rm eq})^2}{(Y_{X_R}^{\rm eq})^2}
Y_{X_R}^2
\right)
\nonumber\\
&\quad
-\langle \sigma v^2 \rangle_{Z',X_R,Y\rightarrow X_R,Y}\left(
Y_{Z'}Y_{X_R}Y_y
-Y_{Z'}^{\rm eq}
Y_{X_R}Y_y
\right)
-\frac{1}{2}\langle \sigma v^2 \rangle_{Z',Y,Y\rightarrow Y,Y}\left(
Y_{Z'}Y_y^2
-Y_{Z'}^{\rm eq}
Y_y^2
\right)
\nonumber\\
&\quad
-3\langle \sigma v^2 \rangle_{Z',Z',Z'\rightarrow Y,Y^*}\left(
Y_{Z'}^3
-\frac{(Y_{Z'}^{\rm eq})^3}{(Y_y^{\rm eq})^2}
Y_y^2
\right)
-\frac{1}{4}\langle \sigma v^2 \rangle_{Z',Y,Y^* \rightarrow Y,Y^*}\left(
Y_{Z'} Y_y^2
-Y_{Z'}^{\rm eq}
Y_y^2
\right)
\nonumber\\
&\quad
-\langle \sigma v^2 \rangle_{Z',X_I,X_R\rightarrow Y,Y^*}\left(
Y_{Z'}Y_{X_I}Y_{X_R}
-\frac{Y_{Z'}^{\rm eq}Y_{X_I}^{\rm eq}Y_{X_R}^{\rm eq}}{(Y_y^{\rm eq})^2}
Y_y^2
\right)
-\langle \sigma v^2 \rangle_{Z',X_R,X_R\rightarrow Y,Y^*}\left(
Y_{Z'}Y_{X_R}^2
-\frac{Y_{Z'}^{\rm eq}(Y_{X_R}^{\rm eq})^2}{(Y_y^{\rm eq})^2}
Y_y^2
\right)
\nonumber\\
&\quad
-\langle \sigma v^2 \rangle_{Z',X_I,X_I \rightarrow Y,Y^*}\left(
Y_{Z'}Y_{X_I}^2
-\frac{Y_{Z'}^{\rm eq}(Y_{X_I}^{\rm eq})^2}{(Y_y^{\rm eq})^2}
Y_y^2
\right)
\Bigg]
\,.
\end{align}
}

\acknowledgments
The work is supported in part by the DFG Collaborative Research Centre “Neutrinos and Dark Matter in Astro- and Particle Physics” under Grant No. SFB 1258 (SMC), by KIAS Individual Grants, Grant No. PG021403 (PK), and by National Research Foundation of Korea (NRF) Grant No. NRF-2019R1A2C3005009 (PK), funded by the Korea government (MSIT), and the Fundamental Research Funds for the Central Universities, by the National Science Foundation of China under Grant No. 11905149 (JL).
The authors thank Chih-Ting Lu and Shu-Yu Ho for mentioning on the $\lambda_{XY}$ term.



\end{document}